\documentclass{aa}  

\usepackage{graphicx}
\usepackage{hyperref}
\usepackage{txfonts}
\usepackage{xcolor}

\begin{document}

   %\title{NF and NNclass}
   \title{Exploring the halo-galaxy connection with probabilistic approaches}

   \titlerunning{The halo-galaxy connection with a probabilistic approach}

    \authorrunning{Rodrigues, de Santi, Abramo \& Montero-Dorta}

   \author{Nat\'alia V. N. Rodrigues\inst{1, 2, 3}\thanks{nvilla-ext@iac.es},
           Natal\'i S. M. de Santi\inst{3, 4}\thanks{natalidesanti@gmail.com},
           L. Raul Abramo\inst{3}
           %\thanks{raulabramo@usp.br}
          \and
          Antonio D. Montero-Dorta\inst{5}%\fnmsep\thanks{Just to show the usage of the elements in the author field}
          }

   \institute{Instituto de Astrof\'{\i}sica de Canarias, s/n, E-38205, La Laguna, Tenerife, Spain
   \and Departamento de Astrof\'{\i}sica, Universidad de La Laguna, E-38206, La Laguna, Tenerife, Spain
   \and Departamento de F\'isica Matem\'atica, Instituto de F\'{\i}sica, Universidade de S\~ao Paulo, Rua do Mat\~ao 1371, CEP 05508-090, S\~ao Paulo, Brazil
   % \\    \email{wuchterl@amok.ast.univie.ac.at}
    \and Center for Computational Astrophysics, Flatiron Institute, 162 5th Avenue, New York, NY, 10010, USA
    \and Departamento de F\'isica, Universidad T\'ecnica Federico Santa Mar\'ia, Avenida Vicu\~na Mackenna 3939, San Joaqu\'in, Santiago, Chile\\
             %\email{c.ptolemy@hipparch.uheaven.space}
             %\thanks{The university of heaven temporarily does not accept e-mails}
             }

   \date{Received October 23, 2024; accepted XXXX XX, XXXX}

% \abstract{}{}{}{}{} 
% 5 {} token are mandatory
 
  \abstract
   {% The connection between galaxies and their host dark matter halos encompasses a range of intricate and interrelated processes, playing a pivotal role in our understanding of galaxy formation and evolution. Traditionally, this link has been established by means of physical or empirical models. Machine learning techniques, on the other hand, are adaptable tools capable of grasping intricate associations between numerous attributes.
   The connection between galaxies and their host dark matter halos encompasses a range of intricate and interrelated processes, playing a pivotal role in our understanding of galaxy formation and evolution. 
   Traditionally, this link has been established through physical or empirical models. On the other hand, machine learning techniques are adaptable tools capable of handling high-dimensional data and grasping associations between numerous attributes.  
   %However, these approaches often lead to deterministic predictors that do not capture the stochasticity inherent to these highly complex processes and relations.
   In particular, probabilistic models in machine learning capture the stochasticity inherent to these highly complex processes and relations.}
   {%We predict the properties of central galaxies based on the properties of their host halos, with different probabilistic approaches which are able to recover the relationships between halo and galaxy properties, as well as the joint distributions of galaxy properties.
   We compare different probabilistic machine learning methods to model the uncertainty in the halo-galaxy connection and efficiently generate galaxy catalogs that faithfully resemble the reference sample by predicting joint distributions of central galaxy properties, namely stellar mass, color, specific star formation rate, and radius, conditioned to their host halo features.}
   {%\Natali{We used galaxies and dark matter halos from the magnetohydrodynamical simulation IllustrisTNG300.
   %The probabilistic approaches we used to recover the galaxy probability density distributions} are based in machine learning methods that model the distribution in different ways: first, a multivariate Gaussian; second, a Categorical distribution where the classes were defined in a high-dimensional parameter space by means of a Voronoi cell-based hierarchical scheme; and third, using the method of normalizing flows.
   %\Natali{All the results are validated considering scores and metrics designed to test probability density distributions}
   The analysis is based on the IllustrisTNG300 magnetohydrodynamical simulation. 
   The machine learning methods model the distributions in different ways. 
   We compare a multilayer perceptron that predicts the parameters of a multivariate Gaussian distribution, a multilayer perceptron classifier, and the method of normalizing flows. The classifier predicts the parameters of a Categorical distribution, which are defined in a high-dimensional parameter space through a Voronoi cell-based hierarchical scheme.
   The results are validated with metrics designed to test probability density distributions and the predictive power of the methods.
   }
   {%\Natali{Four galaxy properties (stellar mass, radius, color and Star Formation Rate) are predicted jointly in a four-dimensional space with high accuracy and prediction.
   %The three methods exhibit similar performances, with Normalizing Flows overcoming neural network classification and Gaussianization in most scenarios.}
   We evaluate the model's performances under various sample selections based on halo properties. The three methods exhibit comparable results, with normalizing flows showing the best performance in most scenarios. The models not only reproduce the main features of galaxy properties distributions with high-fidelity, but can also be used to reproduce the results obtained with traditional, deterministic, estimators.
   Our results also indicate that different halos and galaxy populations are subject to varying degrees of stochasticity, which has relevant implications for studies of large-scale structure.
   }
   {}

   \keywords{halo-galaxy connection --
                machine learning probabilistic methods --
                hydrodynamical simulations
               }

   \maketitle
%
%-------------------------------------------------------------------

\noindent
% {\bf Other title proposals:} 

% \begin{quote}
    
% Exploring the halo-galaxy connection with probabilistic approaches

% The ghost in the machine of the halo-galaxy connection    

% When probabilistic approaches beats deterministic methods for halo-galaxy connection    

% ...?

% Mimicking the halo-galaxy connection using machine learning II: modeling the scatter

% A study of the halo-galaxy connection with machine learning probabilistic models

% \end{quote}

\section{Introduction}

We describe structure formation in terms of a layered approach. At the most basic level we have the density field, in dark matter and in baryons, which along with the gravitational potentials carry all the information about the initial conditions and the physical mechanisms that play a part in the growth of structures. 
At an intermediate level we have the halos and sub-halos, a summary statistic of the underlying density field which express and reflect in a vastly simpler fashion the fundamental cosmological parameters behind structure formation.
And finally, at an observational level we have galaxies, quasars and other tracers of large-scale structure, which are the objects we actually employ when testing the physical models.
It is critical, therefore, to understand how galaxies are related to their host halos and to the larger environment where they formed.

Machine learning has become a part of the standard toolbox used to explore that connection, not only because of its power to replicate the intricate relations between galaxies and dark-matter halos, but also as a means of investigating the physical mechanisms that shape this fundamental link. 
However, the number of factors that determine which properties will emerge for each galaxy are enormous, leading to a degree of stochasticity that has been so far difficult to capture with existing approaches \citep{Wechsler2018, Jo2019, Stiskalek2022, deSanti2022, Jespersen2022, Rodrigues2023, Chuang2024}.

This paper brings important improvements to predictions of galaxy properties based on their dark-matter halos. It is built upon two prior studies in which we examined this relationship through the lens of machine learning, utilizing data from the IllustrisTNG300 simulation \citep{deSanti2022, Rodrigues2023}.
%This paper is a continuation of two previous works where we study the halo-galaxy connection from a ML perspective using the IllustrisTNG300 simulation data.

In the first work of this series \citep{deSanti2022}, we compare several methods, namely neural networks (NNs), extremely randomized trees (ERT), light gradient boosting machines (LGBM) and k-nearest neighbors (kNN), that input a set of halo properties and output the central galaxy properties. We show that these methods are able to reproduce key aspects of the relation between halo and galaxy properties. However, these are all deterministic estimators, which means that for a given halo they are designed to return a single (expected) value for the central galaxy property. 
From a probabilistic standpoint, this is equivalent to focusing the predictions on the peak of the complete distribution, with the consequence of overpredicting the mode of the distribution, and underpredicting the tails. We initially tried to mitigate this problem by treating the problem as being due to an imbalanced dataset, which led us to employ the synthetic minority over-sampling technique for regression with Gaussian noise (SMOGN) \citep{Kunz2019, pmlr-v74-branco17a}.
Although this technique allowed us to better reproduce the halo-galaxy and galaxy-galaxy properties, the improvement was not remarkable, and within that framework we are still constrained to predicting each galaxy property separately.

In the second work \citep{Rodrigues2023}, we concentrate on the same problem, but employing a non-deterministic approach in order to predict probability distributions associated with the galaxy properties, instead of single values. The method consists of binning the continuous galaxy properties and applying a NN classifier, dubbed NNclass, to predict the scores (weights) associated with each of the bins. Importantly, this scores add up to one, providing a proxy for a probability distribution. NNclass works as a proof of concept based on the notion of predicting distributions, instead of single values, in order to properly reproduce the scatter in the halo-galaxy connection. With this approach, one can generate many catalogs of galaxies by sampling from the predicted distribution.
We managed to faithfully reproduce the shape of univariate and bivariate distributions.
However, it is a simple method with key limitations.

First, it is not efficient in terms of predicting multivariate distributions in high dimensional spaces.
In this work, we overcome this problem by introducing the \textit{Hierarchical Allocating Voronoi method} (HiVAl), that allows to apply NNclass to predict the distributions of a larger number of galaxy properties jointly.
There is a variety of ML methods in the literature designed to output distributions instead of single values, which could be used to capture the uncertainty in the relation between halo and galaxy properties. One example is to assume a (multivariate) Gaussian distribution.
However, the distributions of galaxy properties can be very complex, even present multi-modality. Although NNclass allows for some flexibility in the shape of the output distribution with the binning scheme, the method is still constrained to the choice of the binning and the bins can not be arbitrarily narrow, imposing a limitation on the resolution of the distribution.

As opposed to methods where we assume a parametric distribution to describe the target, generative models are designed to learn the underlying distribution of the data itself, in a non-parametric way. In particular, normalizing flows (NF) have proven to be excellent methods to model conditional distributions with complex shapes \citep{JMLR:v22:19-1028}, and even in the context of halo-galaxy connection \citep{Lovell2023}.
%but conditioned to astrophysical and cosmological parameters \citep{Lovell2023}.

In this work we apply machine learning methods to predict joint probability distributions of galaxy properties conditioned to the halo properties. Each method models the target (galaxy) distribution in a different way. We compare a multivariate Gaussian distribution (dubbed NNgauss), a Categorical distribution (NNclass + HiVAl) and the distribution learned with NF. 
The paper is organized as follows. In Section 2 we describe the dataset, the simulation, the halo and galaxy properties considered in this work. In Section 3 we introduce the ML methods. In Section 4 we present the metrics used to quantify the performance and to compare the methods. In Section 5 we present the results and summarize our main conclusions in Section 6.

\section{Data}\label{data}

This work is based on data from the IllustrisTNG magnetohydrodynamical cosmological simulation \citep{Pillepich2018b, Pillepich2018, Nelson2018_ColorBim, Marinacci2018, Naiman2018, Springel2018, Nelson2019}, which was generated using the {\sc arepo} moving-mesh code \citep{Springel2010}. IllustrisTNG (or TNG for simplicity) is an improved version of the previous Illustris simulation \citep{Vogelsberger2014a, Vogelsberger2014b, Genel2014}, featuring a variety of updated sub-grid models that were calibrated to reproduce observational constraints such as the $z=0$ galaxy stellar mass function or the cosmic SFR density, to name but a few (see the aforementioned references for more information). The IllustrisTNG simulation adopts the standard $\Lambda$CDM cosmology \citep{Planck2016}, with parameters $\Omega_{\rm m} = 0.3089$,  $\Omega_{\rm b} = 0.0486$, $\Omega_\Lambda = 0.6911$, $H_0 = 100\,h\, {\rm km\, s^{-1}Mpc^{-1}}$ with $h=0.6774$, $\sigma_8 = 0.8159$, and $n_s = 0.9667$.

Our analysis builds on the works of \cite{deSanti2022} and \cite{Rodrigues2023}, where we employed a variety of ML techniques to reproduce the halo--galaxy connection. In those analyses, we used the largest box available in the TNG suite, TNG300, spanning a side length of $205\,\,h^{-1}$Mpc with periodic boundary conditions. The reason for this choice was to  to minimize cosmic variance for our clustering measurements. For consistency with those works, and in order to allow comparison, we stick to the same dataset in the present work (even though clustering measurements are not presented this time). TNG300 contains 2500$^3$ DM particles of mass $4.0 \times 10^7$ $h^{-1} {\rm M_{\odot}}$ and 2500$^3$ gas cells of mass $7.6 \times 10^6$ $h^{-1} {\rm M_{\odot}}$. 
The adequacy of TNG300 in the context of clustering science has been extensively proven in a variety of analyses (see, e.g., \citealt{Contreras2020, Gu2020, Hadzhiyska2020, MonteroDorta2020, Shi2020, Hadzhiyska2021, MonteroDorta2021A, MonteroDorta2021B, Favole2021, MonteroDorta2024}). 

Also for consistency with \cite{deSanti2022} and \cite{Rodrigues2023}, the same halo and galaxy properties analyzed in those works are employed here. In terms of input halo properties we use:

\begin{itemize}

    \item {\it{Virial mass}} ($M_{\rm vir} [h^{-1} {\rm M_{\odot}}]$), which is computed by adding up the mass of all gas cells and DM particles contained within the virial radius $R_{\rm vir}$ (based on a standard collapse density threshold of $\Lambda_c =200$). In order to ensure that haloes are well resolved, we impose a mass cut $\log_{10}(M_{\rm vir}[h^{-1}{\rm M_{\odot}}]) \geq 10.5$, corresponding to at least 500 dark matter particles.
    
    \item {\it{Virial concentration}} ($c_{\rm vir}$), defined in the standard way as the ratio between the virial radius and the scale radius, i.e.,  $c_{\rm vir} = R_{\rm vir}/R_{\rm s}$. Here, $R_{\rm s}$ is obtained by fitting the DM density profiles of individual haloes with a NFW profile \citep{nfw1997}.

    \item {\it{Halo spin}} ($\lambda_{\rm halo}$), for which we adopt the \cite{Bullock2001_2} definition: $\lambda_{\rm halo} = |J|/\sqrt{2} { M_{\rm vir}} {V_{\rm vir}} { R_{\rm vir}}$. Here, $J$ and $V_{\rm vir}$ correspond to the angular momentum of the halo and its circular velocity at $R_{\rm vir}$, respectively.

    \item {\it{Halo age}}, parametrized as the half-mass formation redshift $z_\text{1/2}$. This age parameter corresponds to the redshift at which half of the present-day halo mass has been accreted into a single subhalo for the first time. The formation redshift is measured following the progenitors of the main branch of the subhalo merger tree computed with {\sc sublink}, which is initialized at $z = 6$. 

    \item The {\it{overdensity}} around haloes on a scale of 3 $h^{-1}$Mpc ($\delta_3$). The overdensity is defined as the number density of subhalos within a sphere of radius $R = 3 h^{-1}$Mpc, normalized by the total number density of subhalos in the TNG300 box \citep[e.g.,][]{Artale2018, Bose2019}.
    
\end{itemize}

As for the galaxy properties that we want to reproduce, we again choose:

\begin{itemize}
    \item The {\it{stellar mass}} ($M_\ast{}$ [$h^{-1} {\rm M_{\odot}}$]), including all stellar cells within the subhalo. In order to ensure that galaxies are well resolved, we impose a mass cut $\log_{10}(M_{\ast{}}[h^{-1} {\rm M_{\odot}}]) \geq 8.75$, corresponding to at least 50 gas cells.

    \item The {\it{colour}} $g - i$, computed from the rest-frame magnitudes, which are obtained in IllustrisTNG by adding up the luminosities of all stellar particles in the subhalo (\citealt{Buser1978}). This specific choice of colour is rather arbitrary. We have checked that using other combinations (i.e., $g - r$) provides similar results. 

    \item The {\it{specific star formation rate}} (sSFR [$ {\rm yr^{-1}} h$]), which is the  star formation rate (SFR) normalised by stellar mass. The SFR is computed by adding up the star formation rates of all gas cells in the subhalo. Note that around $14\%$ of the galaxies at redshift $z=0$ in TNG300 have SFR$=0$. In order to avoid numerical issues, we have adopted the same approach as in \cite{deSanti2022}, assigning to these objects artificial values of SFR, such that they end up distributed around $\rm \log_{10}(sSFR[yr^{-1}h]) = - 13.5$/$\rm \log_{10}(SFR[M_{\odot} yr^{-1}]) = - 3.2$, with a dispersion of $\sigma = 0.5$.
    %The sSFR in our sample covers sSFR $\in [-17.00, -8.30] \rm yr^{-1} 
 
    \item The {\it{galaxy size}}, represented by the stellar (3D) half-mass radius ($R_{1/2}^{(*)} [h^{-1} \, {\rm kpc}]$) -- i.e., %$R_{1/2}^{(*)}$ is 
    the comoving radius containing half of the stellar mass in the subhalo.
    %The galaxy radius in our sample covers $R_{1/2}^{(*)} \in [0.05, 2.00] h^{-1} \, {\rm kpc}$.
\end{itemize}

We refer the reader to the TNG webpage (\href{https://www.tng-project.org}{https://www.tng-project.org}) and/or main papers for more information on the determination of these quantities. Some more details can also be found in \cite{deSanti2022} and \cite{Rodrigues2023}.

%The models are trained with the five halo properties listed above, but they can also be used if there are missing features, i.e., if not all properties are provided (see the github repository for more details).

%--------------------------------------------------------------------

\section{Methods}\label{methods}

In this section we describe the models to predict galaxy properties based on the halo properties. All methods follow the general idea of predicting for each halo a conditioned probability distribution for the properties of its central galaxy. The model's parameters are optimized such that the likelihood of observing the reference catalog (TNG300 training set) is maximized, i.e., the loss function is the negative log-likelihood of the distribution. They differ from each other mainly in the way that the galaxy properties distribution (the likelihood) is modeled. 

\subsection{Neural network with Gaussian likelihood}
In this method we assume a multivariate Gaussian to model the target distribution. We use \href{https://www.tensorflow.org}{\texttt{TensorFlow}} \citep{tensorflow2015-whitepaper} and \href{https://www.tensorflow.org/probability}{\texttt{TensorFlow Probability}} \citep{2017tensorflowprob} to train a multilayer perceptron (MLP) neural network to learn the parameters of the distribution, i.e., the mean values of each galaxy property and the covariance matrix. The Gaussian distribution of an instance characterized by a data vector $\mathbf{x}$ is given by
\begin{equation}
    p(\mathbf{x}|\mathbf{\mu}, \mathbf{\Sigma}) = \frac{1}{\sqrt{(2\pi)^D\text{det}(\Sigma)}}\text{exp}\bigg(-\frac{1}{2}(\mathbf{x} - \mathbf{\mu})^{\intercal}\mathbf{\Sigma}^{-1}(\mathbf{x} - \mathbf{\mu})\bigg)
\end{equation}
where $\mu$ is the vector of mean values whose components are the galaxy properties, $\Sigma$ is the covariance matrix, and $D$ is the number of dimensions (e.g. $D=4$ if we predict the four galaxy properties jointly). The loss-function is then given by the negative log-likelihood of the Gaussian distribution. We refer to this method as NNgauss.

\subsection{Neural network classifier}

This method approximates the target probability distribution as a Categorical distribution whose parameters are the probabilities of the classes\footnote{This is a generalization of the Bernoulli distribution.}. We discretize the continuous values of galaxy properties into several bins that become classes (categories). The probabilities of the classes are computed by training, once again, a MLP using \href{https://www.tensorflow.org}{\texttt{TensorFlow}} and \href{https://www.tensorflow.org/probability}{\texttt{TensorFlow Probability}}. The last layer of the MLP classifier has the softmax activation function, which normalizes the output values in such a way that the scores of all classes add up to one. In this sense, the output values are interpreted as probabilities associated with the classes (the parameters of the Categorical distribution). The loss function is the categorical cross-entropy (CCE), which is the negative log-likelihood of the Categorical distribution:
\begin{equation}
\label{eq:cce}
    {\rm CCE} = -\frac{1}{N}\sum_n^N\sum_k^K y_{nk}\log p_{nk},
\end{equation}
where $N$ is the number of instances, $K$ the number of classes, $y_{nk}$ is the true class and $p_{nk}$ is the assigned probability.

In principle, this strategy is flexible to model distributions with complex shapes, which could handle, e.g., skewness and bimodality observed in the color or (s)SFR of galaxies. The narrower the bins, the higher the resolution, and the better we approach the continuous distribution.
However the bins can not be arbitrarily small, otherwise there would not be a sufficiently large number of instances populating the bins to properly train the MLP.
Therefore, the choice of the number of bins is a balance between the occupancy and the resolution, i.e., the number of objects populating each bin and the size of the bin.

In our precursor work \citep{Rodrigues2023}, we have applied this technique and we managed to reproduce univariate and bivariate distributions of galaxy properties. However, the method was limited to low dimensions because of the way the bins (classes) were defined.
In that paper, we simply grid the distributions in 50 equally spaced bins per dimension, which means that for a three dimensional problem, for example, one would have an order of $10^5$ classes (bins).
In this work, we introduce the \textit{Hierarchical Voronoi Allocation} method (see \S\ref{hival}) to discretize the distributions in such a way that more properties can be predicted jointly.

The strategy to sample from NNclass involves two steps. The MLP outputs the Categorical distribution from which we sample the HiVAl domains (discrete values). Then we sample continuous values from the domains by defining a Normal distribution with mean value corresponding to a randomly chosen object belonging to the domain and variance corresponding to the variance of the domain. The list of objects belonging to the domains and their variances are both HiVAl outputs.

\subsubsection{Hierarchical Voronoi Allocation}\label{hival}

\begin{figure*}
    \centering
    \includegraphics[width=0.245\linewidth]{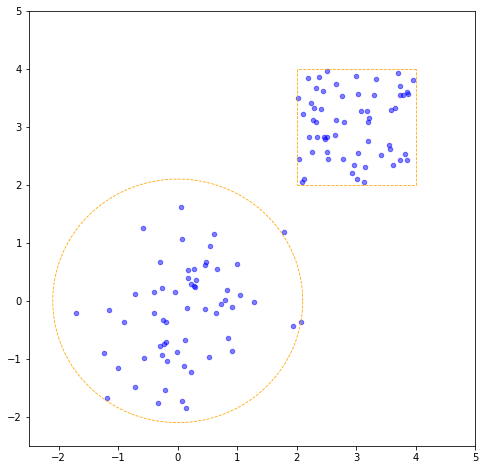}
    \includegraphics[width=0.245\linewidth]{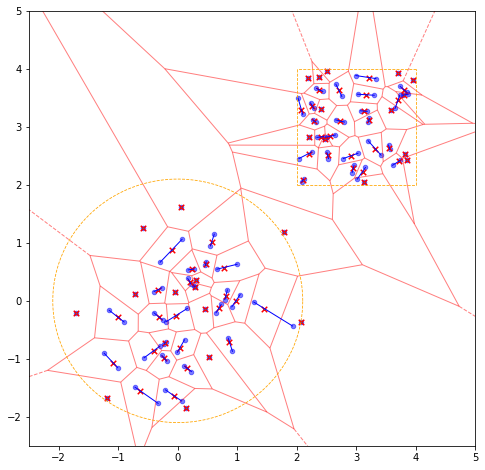}
    \includegraphics[width=0.245\linewidth]{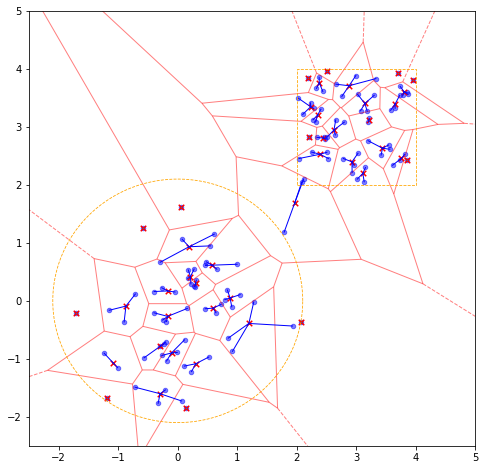}
    \includegraphics[width=0.245\linewidth]{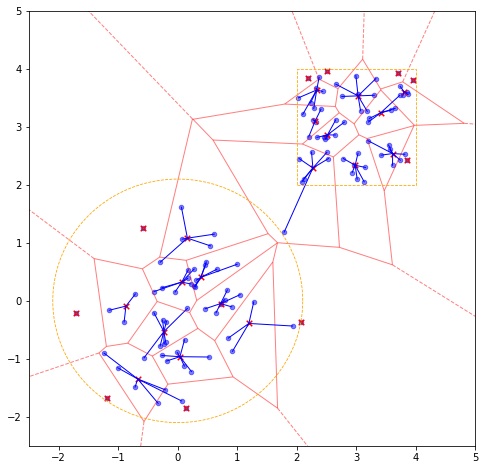}
    \caption{From left to right: original distribution (blue dots, first panel), first iteration of HiVAl (red crosses and boundaries, second panel), second iteration of HiVAl (third panel), and third iteration of HiVAl (fourth panel).
    The yellow lines are simply guides for the boundaries of the two disjoint distributions used in this example. }
    \label{fig:HiVAl}
\end{figure*}

In this section we describe a new tiling and allocation method for collecting discrete objects into cells (or classes), which we call the {\em Hierarchichal Voronoi Allocation} method (HiVAl).
Given a set of $N$ objects in $D$ dimensions, HiVAl works by (i) determining the Voronoi cells of the $N$ objects, and then (ii) pairing neighbouring cells.
In order to determine which neighbours are paired, we employ as criteria the D-dimensional volumes of the Voronoi cells, the (D-1)-dimensional areas of the faces of the cells adjacent to their neighbors, as well as the distances between the neighbouring cells.
Methods based on the Voronoi tesselation have been used before in the context of large-scale structure, e.g., in the void-finding algorithm ZOBOV \citep{ZOBOV}.

HiVAl is a self-organizing, deterministic method to group instances (the objects) into cells, which can be iterated as many times as needed. Each iteration of the method lowers the number of cells by a factor $\gamma \gtrsim 1/2$, and consequently increases the number of objects per cell by a factor $1/\gamma \lesssim 2$.
HiVAl preserves the shape of the original (parent) distribution by progressively tiling the D-dimensional space while preserving the boundaries of the distribution.
The edges of the distribution are frozen by means of identifying the objects which sit at the boundaries (whose Voronoi cells have infinite volumes), and blocking them from pairing up with neighbours.

The way HiVAl works is pictured in Fig. \ref{fig:HiVAl}. 
For that example we took two 2D disjoint distributions: (i) a 2D multivariate Gaussian random distribution with mean $(0,0)$ and diagonal variance $(0.7,0.7)$, and (ii) a 2D uniform random distribution in the interval $(2-4,2-4)$. 
We then sampled 60 objects from each one of the two distributions, for a total of 120 objects -- see the top panel of Fig. \ref{fig:HiVAl}. 
The 3$\sigma$ region corresponding to the 2D Gaussian distribution is shown in that figure as the yellow circle, and the boundary of the uniform distribution is shown as the yellow square.

In the first iteration of HiVAl (second panel in Fig. \ref{fig:HiVAl}), for each cell (in this case, for each object), the algorithm determines the Voronoi domains of the objects and identifies the neighbouring cells.
If the object sits at the boundary of the distribution, its Voronoi cell volume is infinite and it is therefore blocked from pairing up with a neighbour.
But if the cell has a finite volume, and if it has available finite-volume neighbours to pair up with, then the partner is chosen as the neighbour with the largest area in common with the original object.
This choice ensures that the cells created by the resulting pairs have the smallest area/volume ratio. 
The red crosses indicate the ``center of mass'' (CM) of the resulting cells after this first iteration: some are boundary cells (which are the unpaired particles at the boundaries), some sit between pairs of the original objects, and some cells may be single objects simply because there are no available partners anymore.
For this example we have used ``masses'' (weights) of the CM which are inversely proportional to the size of the Voronoi cells, $m_i = V_i^{-1/D}$, but these weights can be tuned to the needs of the particular problem at hand.

In the second iteration of HiVAl (third panel in Fig. \ref{fig:HiVAl}), the algorithm again starts by determining the Voronoi cells and identifying all the neighbours.
Cells with infinite volume are blocked from pairing up with a neighbour, which preserves the boundaries of the original distribution.
The finite-volume cells with available partners are then paired with the neighbour that has the largest area in common with that cell.
The CM (red crosses) now indicate cells that can be made up of 1-4 particles.
The third iteration (see fourth panel of Fig. \ref{fig:HiVAl}) continues this process: the cells are now domains with 1-8 objects.
In this example, after the third iteration we end up with 9 boundary domains (made up of single particles) and 18 domains with 3 to 8 objects.

We fit HiVAl with the same training set used to train the machine learning models (see \S\ref{training procedure}). HiVAl returns, for each iteration, the position of the center of the domains (in parameter space coordinates), the list of training instances with the index of the domain they belong to, i.e, their classes -- which is used as target to train NNclass --, and, complementary, the list of training set instances belonging to each domain, as well as their standard deviation, which characterize the dispersion of the domains. Several complementary scores are provided for additional analysis.

HiVAl approximates the density of the distribution of the data and, therefore, its probability distribution. 
It can be used to analyze the statistical significance of a given region (domain) in the parameter space by evaluating how it is populated, e.g., if it has low or high density.
For the purpose of predicting galaxy properties, we use HiVAl to model their distribution and combine it with NNclass to assign the scores associated to each HiVAl cell.

HiVAl is implemented with \href{https://scipy.org}{\texttt{SciPy}} library \citep{2020SciPy-NMeth}. 
%The code for the example showed in this section can be found in the \texttt{github} repository. \textbf{[CITE github repo here later!]}

%{\color{red} To do/include in the paper:

%* Tests of HiVAl \Natali{[Natalí: Do you think it is still needed? I mean, include more tests in this section?]}

%* HiVAl can be employed to... \Natali{[Natalí: It should be interesting to include tomething around these lines in the paragraph above.]}

%* HiVAl -> GitHub \Natali{[Natalí: I believe Raul is responsible for the code to generate Fig. \ref{fig:HiVAl}. Can you please send it to us? Natália, are you going to create the github repo? I can do this as well, if you prefer. Or help reviewing it as well.]}}

% Para não confundir ninguém:
%Notice that HiVAl is a method completely independent of NNclass. It is a technique that can be used to approximate a target distribution, so in some sense it has the same purpose as a ML generative model.

\subsection{Normalizing flows}

NF is a ML algorithm designed to be a flexible way to learn distributions with NNs.
It is a powerful generative method because it can handle conditional probabilities \citep{bingham2019pyro} in high dimensions (still taking into account the curse of dimensionality \citep{Coccaro2023}).
% It is a powerful generative method because this can be done for distributions in $N$-dimensions (still taking into account the
% curse of dimensionality \citep{Coccaro2023}) and can deal with conditional probabilities \citep{bingham2019pyro}.

% The basic idea of NF is the change of the variable formula for PDFs, transforming a simple distribution (e.g. a Gaussian distribution) into the desired distribution. Mathematically, given two variables $x, z \in \mathbb{R}^D$, where $D$ is the number of dimensions, with PDFs $p_x (x), p_z (z): \mathbb{R}^D \rightarrow \mathbb{R}$, we can define the {\em flow} as a bijective map $f: x \rightarrow z$, which must be invertible $g = f^{- 1}$ and differentiable.
% Under this condition, the two densities are related by:
% \begin{align}
%   p_z (z) & = p_x \left[ f^{- 1} (z) \right] \left| \text{det} \left( \frac{\partial f^{- 1}}{\partial x} \right) \right|
%             = p_x \left[ g (z) \right] \left| \text{det} \left( \frac{\partial g}{\partial z} \right) \right|^{- 1} ,
% \end{align}
% where $\left| \text{det} \left( \frac{\partial f^{- 1}}{\partial x} \right) \right| = J_{ f^{ - 1}} (x)$ is the Jacobian of $f^{- 1} (x)$. 
%Therefore, using the chosen $f$, we can transform the distribution on $x$ and use NF to sample from $p_x (x)$.
The basic idea of NF is to use the change of variables formula for PDFs to transform a simple distribution (e.g. a Gaussian distribution) into the desired distribution. 
Mathematically, suppose we have two variables $x, z \in \mathbb{R}^D$, with PDFs $p_x (x), p_z (z): \mathbb{R}^D \rightarrow \mathbb{R}$, where $D$ is the number of dimensions. 
We can define the {\em flow} as a bijective map $f(z) = x$, $f^{- 1}(x) = z$, which must be invertible and differentiable. 
Then, the two densities are related by:
\begin{align}
  p_x (x) & = p_z \left( z \right) \left| \text{det} \left( \frac{\partial f(z)}{\partial z} \right) \right|^{-1} = p_z \left( f^{- 1} (x) \right) \left| \text{det} \left( \frac{\partial f^{- 1}}{\partial x} \right) \right| ,
\end{align}
where $\left| \text{det} \left( \frac{\partial f^{- 1}}{\partial x} \right) \right| = J_{ f^{ - 1}} (x)$ is the Jacobian of $f^{- 1} (x)$. We choose $p_z(z)$ to be a standard Normal distribution. Once we have $f$, we can easily sample from the base distribution $p_z(z)$ and apply $f$ to obtain a sample from the target distribution $p_x(x)$.

%Due to the properties obeyed by these distributions we can also compose $n$ transformations
%$f_1 \cdot f_2 \cdot \dots \cdot f_n$ in order to increase the flexibility to re-obtain the distribution $p_x (x)$.

One way to learn the bijector that models the distributions with NFs is to use {\em coupling flows} with the so called {\em coupling layers}. A distribution $p_x (x)$ of dimension $D$ is split into two parts $A$ and $B$, with dimensions $d$ and $(D - d)$. Then, the parameters of $A$ are modeled by a NN that uses $B$ as input. Starting with $x \in \mathbb{R}^D$, we can split the variable into $(x^A, x^B) \in \mathbb{R}^d \times \mathbb{R}^{(D - d)}$. Then, the variables transform as
\begin{align}
  x^A & = h \left[ z^A; \Theta \left( z^B \right) \right] \\
  x^B & = g(z^B),
\end{align}
where $g$ and $h$ are both bijections. The parameters of the coupling function $h$ are defined on $\mathbb{R}^{D - d}$ by the generic function $\Theta \left( z^B \right)$, which is modeled by a NN. The inverse transformation is given by
\begin{align}
 z^A & = h^{- 1} \left[ x^A; \Theta \left( z^B \right) \right] \\
 z^B & = g^{-1}(x^B) .
\end{align}
Examples of this model are NICE \citep{NICE2014}, Real NVP \citep{RealNVP2016}, and neural splines, or spline coupling \citep{NeuroSplines2019, Dolatabadi2020}.
More specifically, in the case of spline coupling, $g$ and $h$ are both monotonic rational splines.
%A {\em spline} is a rational of spline bijections of linear and quadratic order. 

Splines are functions designed to build an interpolation of a set of points based on the ratio of $2$ $d$-dimensional polynomials $\alpha^{(k)}$ and $\beta^{(k)}$ with $k$ segments (knots)
\begin{equation}
  y_d = \frac{\alpha^{(k)} (x_d)}{\beta^{(k)} (x_d)} .
\end{equation}
The spline is built on a specific interval $[- K, K]$.

A generalization to learn distributions in NFs is done by the {\em autoregressive flows} \citep{auto_regressive-2016}.
This means that we split the data variable in more dimensions and the transformation for each dimension $i$ is modeled by an autoregressive NN, resulting in a conditional probability distribution given by $p (x_i|x_1, \dots, x_{i - 1})$.
% After each autoregressive layer, the dimensions are permuted, to ensure the expressivity of the bijections over the full dimensionality of the target distribution.
More specifically, considering the {\em coupling function}, or bijector $h (\cdot, \theta): \mathbb{R} \rightarrow \mathbb{R}$,
parametrized by $\theta$, we can define the {\em autoregressive flow} such that 
\begin{align}
  x_1 & = z_1\\
  x_i & = h \left[ z_i; \Theta_i (x_i, \dots, x_{i - 1}) \right], i = 2, \dots, D .
\end{align}
Alternatively, the inverse is
\begin{align}
 z_1 & = x_1\\
 z_i & = h^{- 1} \left[ x_i; \Theta_i (z_1, \dots, z_{i - 1}) \right], i = 2, \dots, D .
\end{align}

% In the case of joint distributions $p_{x, y} ( x, y )$, this can be done by NF as a decomposition of the product of a conditional $p_{y | x} ( y | x )$ and a
% univariate distribution $p_{x} ( x )$ as
As opposed to NNgauss and NNclass, NF is a complete generative method in the sense that it can model the joint distribution of all the variables in the problem (in our case, halo and galaxy properties), as
\begin{equation}
\label{eq:joint}
 p_{x, y} ( x, y ) = p_{y | x} ( y | x) p_{x} ( x ).
\end{equation}
The loss function is then the negative log-likelihood: $-\frac{1}{N}\sum_n^N (\log{p_{y | x} (y | x)} + \log{p_{x} (x)})$.

For our purposes, we only use the conditioned distribution $p(y={\rm gal.}| x={\rm halos})$ to get the probability distribution of the galaxies hosting some set of given halos, e.g., the halos from our test set. Nevertheless, we are also modeling the halo properties distribution itself, following Eq.\eqref{eq:joint}, and we can use $p(x = \text{halos})$ to handle missing features (see the github repository for more details).

In the present work we use {\em spline coupling} to learn the halo distribution $p(x = {\rm halo})$, and {\em conditional auto-regressive spline} to learn the conditioned galaxy distribution $p(y = {\rm gal.}|x = {\rm halo})$. We implement NFs using the
\href{https://docs.pyro.ai/en/stable/index.html}{\texttt{Pyro}} library \citep{bingham2019pyro}.

\section{Training procedure}\label{training procedure}

The ML methods have been trained, validated and tested considering the complete dataset of 174,527 halos/galaxies (see \S\ref{data}). We have split the initial catalog into training, validation and test subsets of $48\%$, $12\%$, and $40\%$, respectively.
We use \href{https://optuna.org}{\textsc{Optuna}} package \citep{optuna2019} to perform a Bayesian optimization with Tree Parzen Estimator (TPE) \citep{Bergstra2011} to select the hyperparameters for each of the different models. We do at least 100 trials for each model.
In the following subsections we describe the set of explored hyperparameters.

\subsection{NNgauss model selection}

The explored hyperparameters are the number of layers and, for each layer, the number of neurons and the parameter for the L2 regularization. We also try different values for the initial learning rate of the Adam optimizer \citep{Adam} and implement a schedule to reduce it when the validation loss stagnates for 20 epochs. We implement an early stopping schedule with a patience of 40 epochs.
% na verdade eu fixei o l2 para todas as camadas :P
The final, best set of hyperparemeters is the one yielding the lowest loss function on the validation set.

\subsection{NNclass model selection}

We explore the same hyperparameters as NNgauss.
One important additional aspect of NNclass model selection is the choice of HiVAl iteration, i.e., the number of classes (bins). This is related with the balance between resolution and occupation. Narrow cells lead to better resolution, but the higher the number of cells, the lower their mean occupancy, and the harder it is for the MLP to properly classify the objects.

% We set the number of classes as a hyperparameter to be explored by \textsc{Optuna}, but we exclude iterations with extremely low occupancy (typically with number of classes greater than 10000) and with extremely low resolution (when even a perfect classification leads to poor correlation between the true and sampled values).

% Since we are comparing different likelihoods when we change the number of bins, we can not rely on the value of the loss function to select the best model. Instead, we compute the PCC of a random catalog generated with the model and choose the one leading to highest PCC (sum of the PCC in all dimensions).

In principle, the number of classes can be set as a hyperparameter to be explored by \textsc{Optuna}. However, we could not rely on the value of the loss function to select the best model because we would be comparing different likelihoods. Moreover, choosing a different metric that grasps information from the $D$-dimensional space may not be trivial.
Alternatively, we choose a few number of classes based on some prior knowledge on occupancy and resolution (i.e., not too large neither too low number of classes), and run \textsc{Optuna} on them independently. Afterwards, we compare the generated samples of continuous values and choose the one with best performance in terms of the metrics discussed in \S\ref{metrics}, giving preference to the model with best conditioning power.

\subsection{NF model selection}

%The explored hyperparameters are number of bins, which are taken by the transformations, both on spline coupling and conditional auto-regressive spline, the split dimension taken on spline coupling, the number of layers taken on conditional auto-regressive spline); the number of hidden features (taken on both transformations); the learning rate (taken in the Adam optimizer \citep{Adam}); and the batch size.

The explored hyperparameters are the following. The number of bins in the spline coupling and in the conditional auto-regressive spline, the split dimension on the spline coupling, the number of layers in the conditional auto-regressive spline, the number of hidden features (neurons) in both transformations (splines), the learning rate in the Adam optimizer and the batch size. We consider the final, best set, of hyperparameters the one that yields the lowest loss function on the validation set.

\section{Metrics}\label{metrics}

In this work we train models to predict probability distributions of galaxy properties based on halo properties to quantify the uncertainty in this relation.
As a consequence, we expect to generate catalogs of galaxy properties that faithfully reproduce the reference by sampling from these distributions. Therefore, we select metrics to validate and measure the quality of the predictions by quantifying the similarity between the distributions, the calibration of the predicted distributions, and the conditioning power of the models to ensure that the right galaxies are populating the right halos.

\subsection{Kolmogorov-Smirnov test}
\label{sec:KStest}

The Kolmogorov-Smirnov test (hereafter K-S test) quantifies the difference between two distributions [$f_1 (x_1)$,
$f_2 (x_2)$] in a non-parametric way by measuring the
maximum distance between the cumulative distribution functions (CDFs)
\begin{equation}
 D = \text{max} \left( |F_1 (x_1) - F_2 (x_2)| \right),
 \label{eq:ks-test}
\end{equation}
where $F_1 (x_1) =$ CDF $[f_1 (x_1)]$, $F_2 (x_2) =$ CDF $[f_2 (x_2)]$.
Similarly, one can measure the 2D K-S test for bivariate distributions (\citealt{ivezic2014statistics, 2DKS-Peacock1983, 2DKS-Fasano1987}).

The K-S test is sensitive to location, scale, and the shape of the underlying distributions, but it is more sensitive to differences near the center than in the tails. It is often used for hypothesis testing to check whether two samples comes from the same distribution and focuses on the maximum difference between the CDFs, making it less sensitive to subtle details.

In this work, we have implemented our own 1D K-S test schedule and used the \cite{2DKS} repository to compute the 2D K-S test.

% When considering not only one, but two independent variables, it may also be useful to compute the 2D K-S test.
% The main idea is the same as for the 1D K-S test, but for two-dimensional data.
% More details can be found in \citealt{ivezic2014statistics, 2DKS-Peacock1983, 2DKS-Fasano1987}.

%The K-S test is sensitive to location, scale, and to the shape of the underlying distributions. It is more sensitive to differences near the center than in the tails, i.e., it is not very sensitive to details in the differential distribution function.
%However, it is not very sensitive to details in the differential distribution function, i.e., it is more sensitive near the center of the distribution than at the tails.
% Good similarities of distributions happen for values close to zero.

%In order to evaluate this metric we have sorted one sample $x_i$, with $i \in [0, 1000]$, of each predicted posterior $\hat{p} (\theta_i | x_i)$ and computed the metric. Also, we averaged the values found for each distribution and we present their standard deviations, related to the $1000$ samples, as the error bars while presenting this metric for each galaxy property and for them predicted in two dimensions.

\subsection{Wasserstein distance}

The Wasserstein distance, sometimes called the earth mover's distance, is a measure of the distance between two probability distributions.
Intuitively, it can be interpreted as the minimum cost of modifying one distribution to obtain another one. Mathematically, given two probability density distributions [$u, v$] this can be written as
\begin{equation}
 W (u, v) = \mathrm{inf}_{\pi \in \Gamma (u, v)} \int_{\mathcal{R} \times \mathcal{R}} d \pi (x, y) |x - y|
\end{equation}
where $\Gamma (u, v)$ is the set of probability distributions on $\mathbb{R} \times \mathbb{R}$ with marginals $u$ and $v$ \citep{Wasserstein2015, 2020SciPy-NMeth}.

The Wasserstein distance provides a metric that takes into account the entire structure of the distribution, allowing for a meaningful and smooth representation of the similarities and differences between them. The Wasserstein distance is an important metric in the context of optimal transport and have been widely used to train generative models, like Wasserstein GANs (generative adversarial networks \citep{arjovsky2017wassersteingan}).
% i.e., it changes continuously with changes in the underlying distribution. 
%It has been used by some works in the literature as loss function, e.g., for GANs.

%Due to the fact that this distance is done along many small portions of the marginals, we have a metric which gives a meaningful and smooth representation of the similarities, and distances, between the complete distributions.

%Some works in the literature uses the Wasserstein distance to compute the loss function to approximate a target distribution.

We compute the 1D Wasserstein distance based on \cite{2020SciPy-NMeth} and the 2D version using \cite{flamary2021pot}.
%We have calculated the Wasserstein distance in the same fashion way as explained at the end of the K-S test subsection \ref{sec:KStest}.

\subsection{Coverage test}

The coverage test is a metric to verify the accuracy of generative posterior estimators $\hat{p} (\theta | x)$, for a high number of dimensions of $\theta$, using coverage checks and is often used in the context of simulation based inference (SBI).
Here we use the \textit{Tests of Accuracy with Random Points} (TARP)\footnote{See the code used at\
 \href{https://github.com/Ciela-Institute/tarp}{https://github.com/Ciela-Institute/tarp}.} \citep{lemos2023}.

At this point, it is useful to explicitly frame our problem using SBI language for clarity. A simulation is a pair $\{\theta_i, x_i\}$, which in our case are the pairs of central galaxies and host halos given by TNG300. The posterior is the probability of observing $\theta$ given $x$, which in our case is modeled by the methods presented in \S\ref{methods}. For each simulation (halo) $i$, we can sample many values $\theta_i$ (galaxy properties) from this estimator and compare with the true value $\theta^\star_i$ from TNG300. Depending on how the samples compare with the true value, according to the coverage test, we can claim that the estimator $\hat{p}(\theta|x)$ is well calibrated.

The basic idea of this score is to compute the {\em expected coverage probability} ECP $(\hat{p}, \alpha, \mathcal{G})$ averaged over the data distribution $x$, where $\hat{p}$ is the posterior estimator, $\mathcal{G}$ is a credible region generator, and $\alpha$ is a {\em credibility level}. ECP is computed using samples from the true joint distribution $p (\theta | x)$ and the samples from the estimated posterior distribution $\hat{p} (\theta | x)$.
% of data $x$ and the parameters of interest $\theta$

More specifically, for each true value $\theta^{\star}_i$, it takes the posterior estimation $\hat{p}_i (\theta_i | x^{\star})$ and sample a reference point $\theta^r_i$\footnote{Not to be confused with the true value $\theta^\star_i$ from TNG300.} from an Uniform distribution $\mathcal{U} (0, 1)$ -- all the samples are transformed to have values within $0$ and $1$.
Then, it measures the distance of each $\theta_i$ to $\theta^r_i$ as $d(\theta_i, \theta^r_i)$ and count the number of samples that are inside a radius $R_i$ defined by the distance between $\theta^{\star}_i$ and $\theta^r_i$, i.e., for each simulation $i$, it computes the fraction $f_i$ of the sampled points falling within a ball centered at $\theta^r_i$.
%summing up those which are inside a radius $R_i$ defined by the distance between $\theta^{\star}_i$ and $\theta^r_i$.  
The ECP is then defined as the average of $f_i < 1 - \alpha$ over the simulations (data points), for credibility levels ranging from 0 to 1.
%This is done until each of these fractions $f_i$ is lower than the credibility level $(1 - \alpha_i)$ for the point $i$ in question, which gives the ECP (coverage) in terms of the credibility level.

A good quality posterior distribution for the whole range of $\theta$ gives a one-to-one correspondence of coverage to credibility level. Deviations from the one-to-one correspondence indicates that the posterior estimator is underfitting, overfitting, or biased. The algorithm is fully described in \cite{lemos2023}.

%In our case, we have estimators for the probability of galaxy properties conditioned to halo properties $\hat{p}(g | h)$ and we want to use the coverage test to quantify the accuracy of these estimators based on the true values from TNG300 and on samples of galaxy properties generated by the estimators.

\subsection{Simulation based calibration}\label{sbc score}
Similarly to the TARP coverage test described above, one can check whether the predicted distributions are well calibrated with the simulation based calibration (SBC) rank statistic \citep{sbc}. Given a halo, we generate samples for the hosted galaxy properties and rank them with respect to the corresponding reference value (from TNG300), i.e., we count how many values fall below the reference. When analyzing many halos, we expect this statistic to follow a Uniform distribution, meaning that the distribution is neither too narrow (overconfident), too broad (underconfident), or biased. The results are presented in Appendix \ref{add. results}.
% We emphasize, however, that this test is a necessary but not sufficient condition to ensure the reliability of the prediction, since it does not inform about the predictive power of the model.

\subsection{Pearson correlation coefficient}
%We use the Pearson correlation coefficient as a metric for maximum likelihood estimation. It is defined as
We use the Pearson correlation coefficient (PCC)
\begin{equation}
    \rm PCC = \frac{\rm cov(y^{pred.}, y^{true})}{\sigma_{y^{pred.}}\sigma_{y^{true}}}
\end{equation}
to quantify the correlation between the predictions and the reference sample and, therefore, evaluate how well the galaxy properties are conditioned to the halo properties, i.e., evaluate the predictive power of model. This metric is particularly informative for single-point estimation as we can verify if the expected values are in agreement with the reference.

\section{Results}\label{results}

In this section we compare the best \textsc{Optuna} trials of NNgauss, NNclass and NF.
We start \S\ref{sec:complete} by comparing the predictions in the complete test set and then, in \S\ref{halo populations}, we compare the model predictions for some halo populations.

Here we show the results computed based on the prediction of the four dimensional distribution of all galaxies properties predicted jointly. However, one could train the models to predict distributions with whatever desired combination of galaxy properties, e.g., single properties, pairs of properties, etc. The advantage of predicting the joint distribution is that it captures correlations between galaxy properties \citep{Rodrigues2023}.

Unless noted otherwise, all metrics are computed with the complete test set and 1000 catalogs randomly sampled from the predicted distributions for each instance (halo/galaxy) to calculate the standard deviation. The exceptions are the 2D K-S test and 2D Wasserstein distance, which are computed using a randomly drawn subset of $30 \times 10^4$ from the test catalog, and 5 samples per instance to calculate the standard deviation. 

Additional results are presented in \ref{add. results}, such as the simulation based calibration analysis introduced in \S\ref{sbc score} and the Tables with the explicit values for the metrics (PCC, K-S test and Wasserstein distance).

\subsection{Complete test sample}
\label{sec:complete}

\begin{figure*}%[!h]
    \centering
    \includegraphics[width=0.35\linewidth]{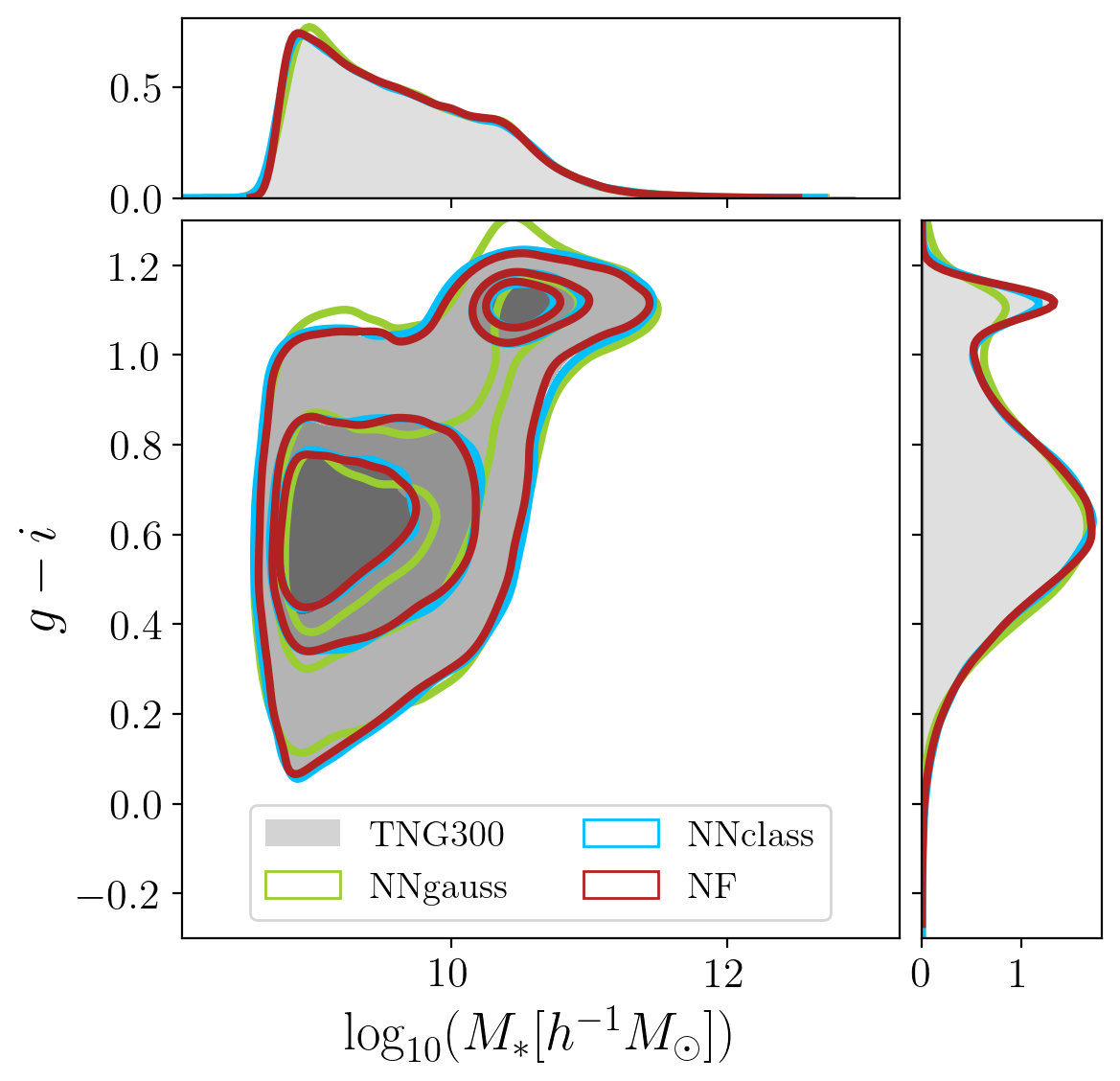}
    \hspace{0.5cm}
    \includegraphics[width=0.35\linewidth]{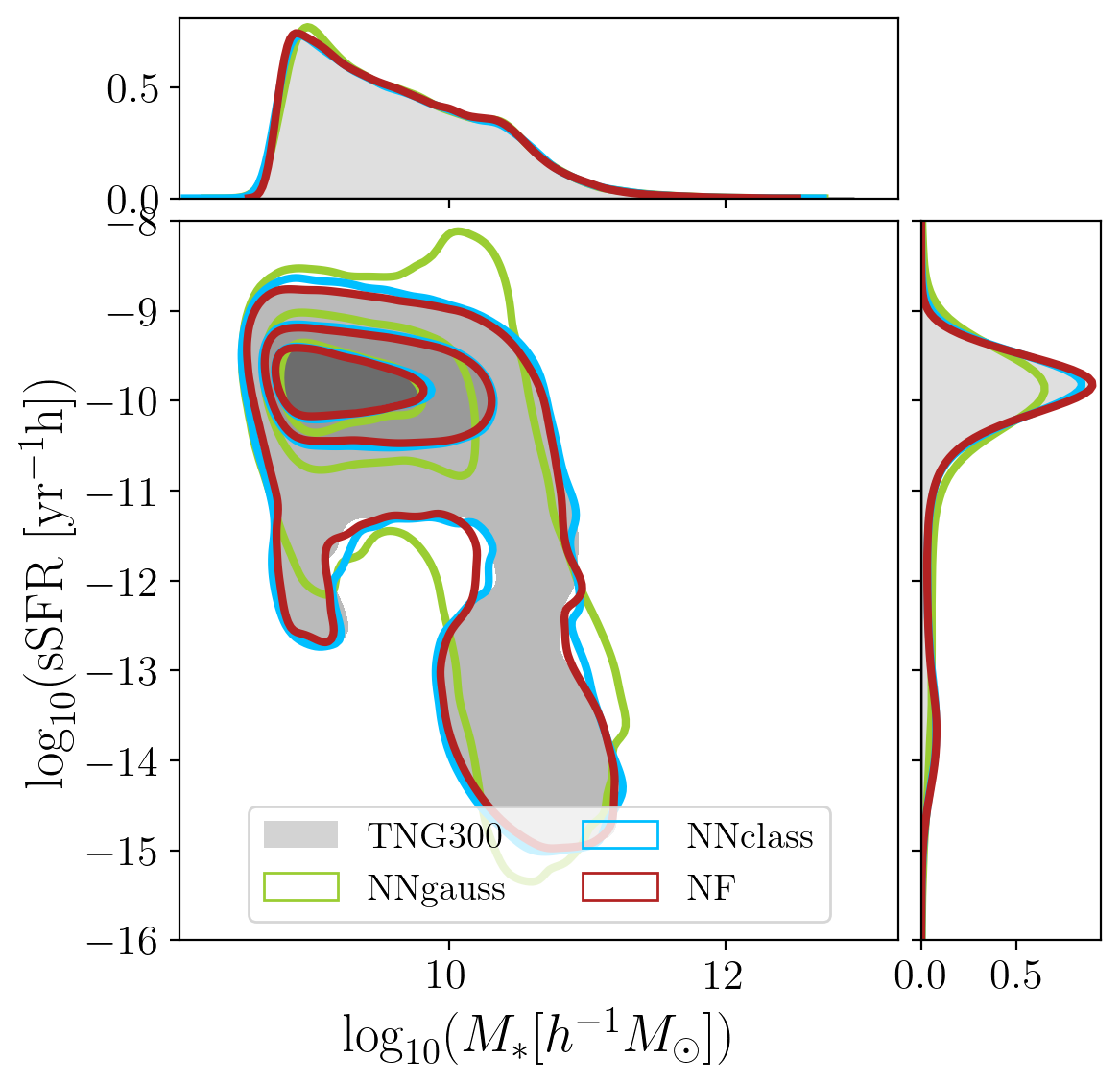} \\
    \includegraphics[width=0.35\linewidth]{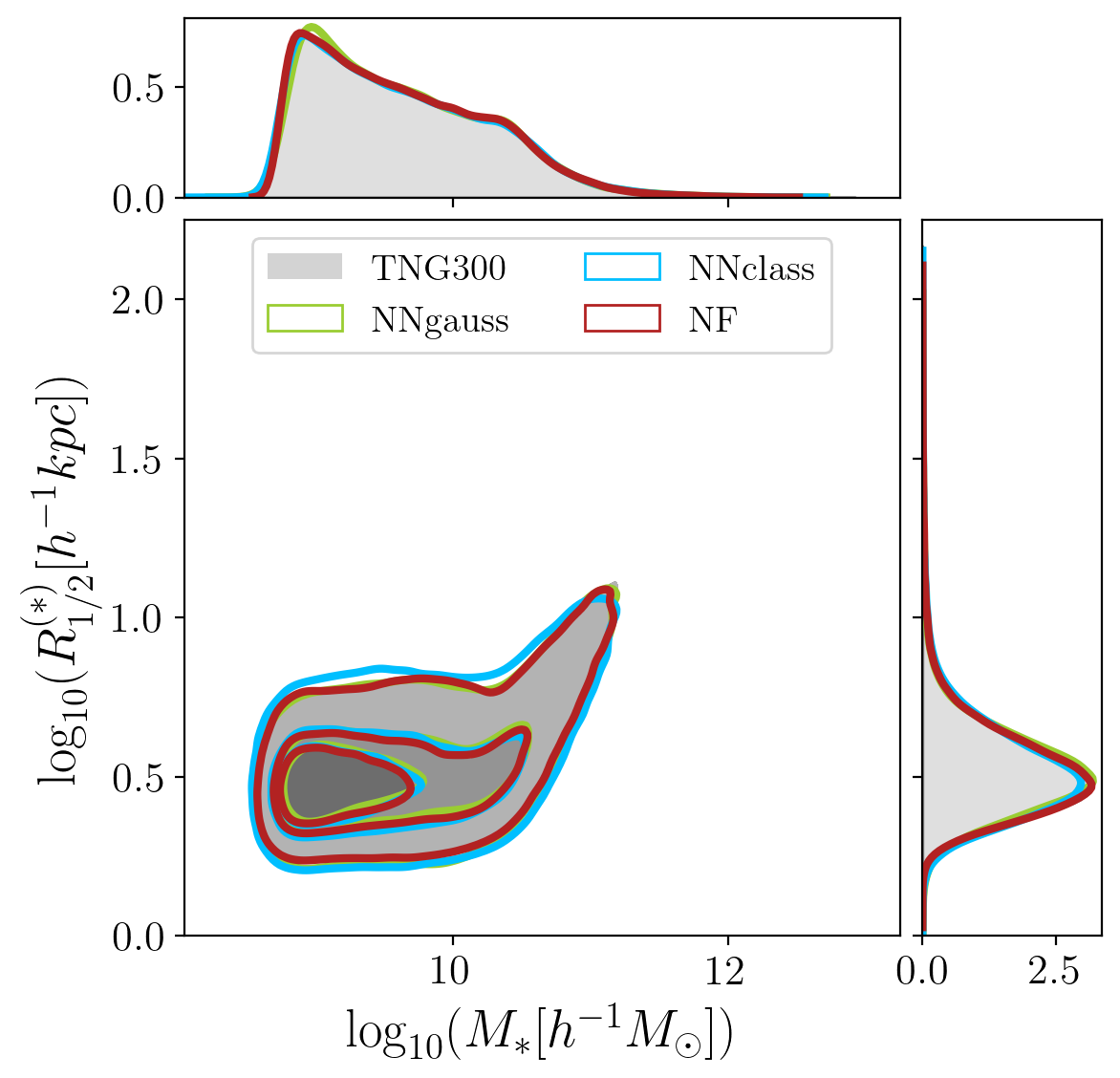}
    \hspace{0.5cm}
    \includegraphics[width=0.35\linewidth]{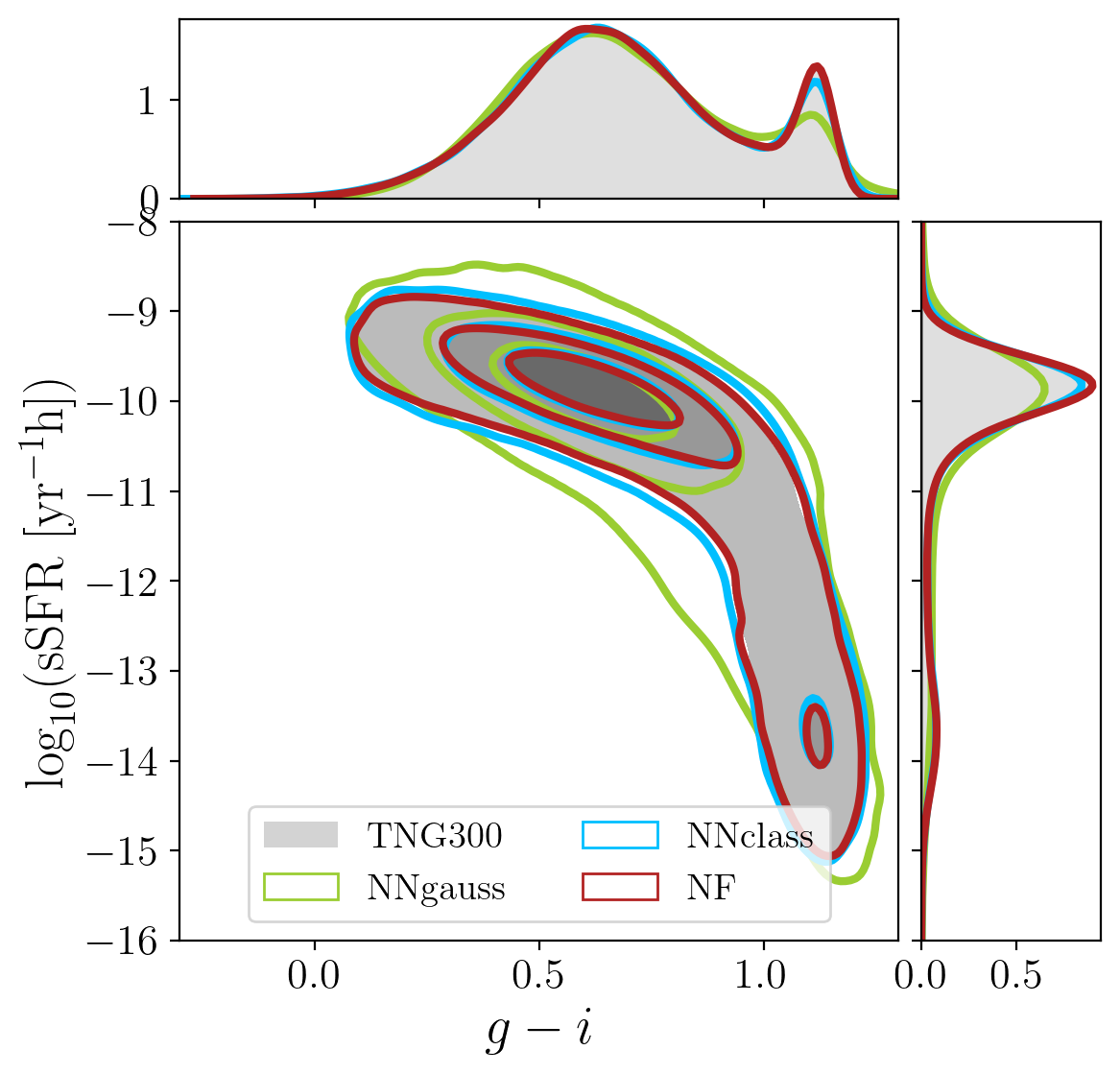}
    \caption{Contour plots of pairs of galaxy properties predicted by each model (NNGauss, NNclass, and NF) compared with the truth distribution of the test set from TNG300. The samples are generated by marginalizing the complete four-dimensional distribution.}
    \label{fig:complete_sample scatter}
\end{figure*}

This section presents the results computed with the complete test sample.
We start with a visual inspection of contour plots of a single generated catalog in Fig.\ref{fig:complete_sample scatter}. It shows the diagrams that are typically analyzed in the context of astrophysics, but one could plot any of the possible combinations of the four galaxy properties by marginalizing over the remaining ones. We see that NNclass and NF faithfully reproduce the shape of the target distributions, both in one and two dimensions, proving their flexibility to model complex shapes. NNgauss shows good agreement for stellar mass and radius, but shows biased contours for the color and sSFR (especially for $3$-$\sigma$), which have a strong bimodal feature.

%\begin{figure*}%[!h]
%    \centering
%    \includegraphics[width=0.245\linewidth]%{complete_sample_smass_color_contour2D.png}
%    \includegraphics[width=0.245\linewidth]{complete_sample_smass_sSFR_contour2D.png}
%    \includegraphics[width=0.245\linewidth]{complete_sample_smass_radius_contour2D.png}
%    \includegraphics[width=0.245\linewidth]{complete_sample_color_sSFR_contour2D.png}
%    \caption{Contour plots of pairs of galaxy properties predicted by each model (NNGauss, NNclass, and NF) compared with the truth distribution of the test set from TNG300. The samples are generated by marginalizing the complete four-dimensional distribution.}
%    \label{fig:complete_sample scatter}
%\end{figure*}

\begin{figure*}
    \centering
    \includegraphics[scale=0.6]{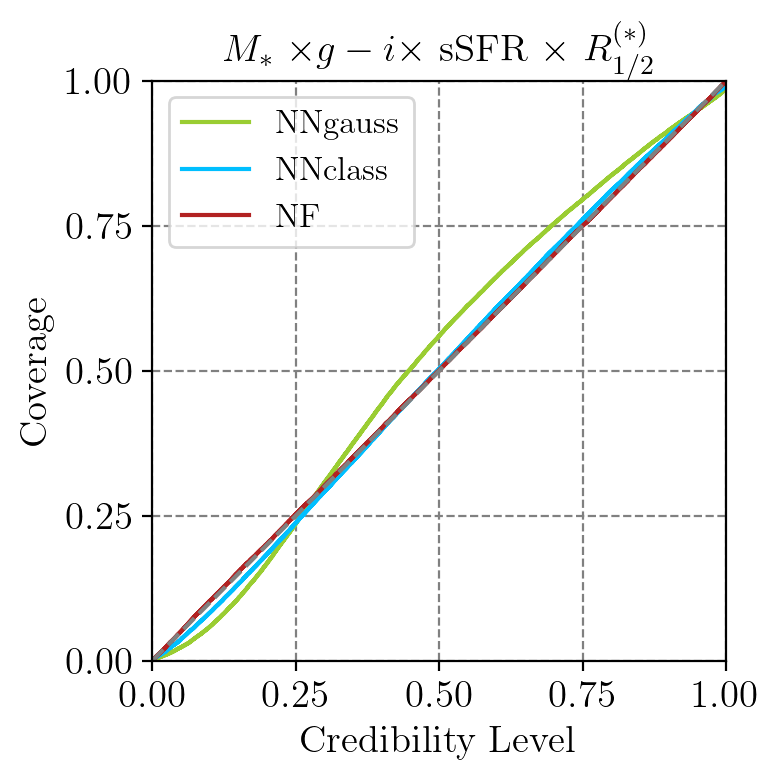}
    \includegraphics[scale=0.6]{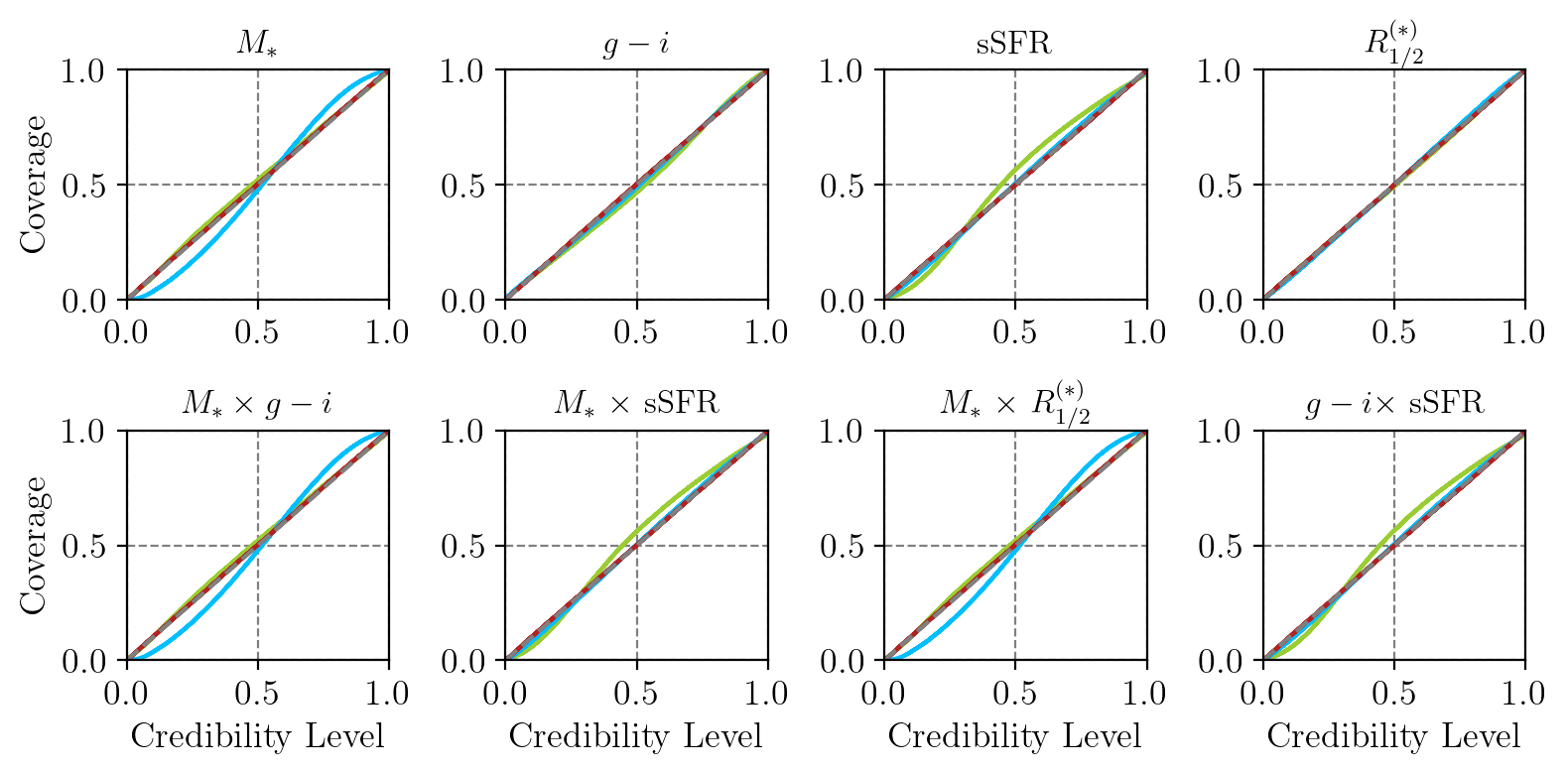}
    \caption{\textit{Left:} TARP coverage test of the joint distribution of all properties. \textit{Right:} TARP coverage test over marginal distributions; top panels show individual properties and bottom panels shows pairs of properties.}
    \label{fig:comple_sample CT}
\end{figure*}

Fig.\ref{fig:comple_sample CT} shows the coverage test of the joint distribution of all four properties in the left-hand side, and, once again, of univariate and bivariate marginalized distributions in the right-hand side. We show the same pairs of properties as Fig.\ref{fig:complete_sample scatter}.
We compute the coverage test with 1000 samples for each instance in the test set\footnote{In the case of NNclass, we sample 40 classes and then 25 continuous values for each classes.}.
Typically, all methods present well calibrated distributions. The main exceptions are NNclass' stellar mass prediction, which indicates that it is underfitting, and NNgauss' sSFR prediction. We verified that NNclass presents this underfitting feature even when the stellar mass is predicted alone. 
%and it worsens as we increase the number of galaxy properties predicted jointly. This indicates a limitation of the method that could perhaps be improved with a different parametrization of stellar mass when discretizing this property.

The biased prediction of NNgauss is likely due to the bimodal shape in sSFR. Notice that for $g - i$ it also slightly deviates from the diagonal. This could perhaps be improved with a mixture of Gaussian distributions, e.g., Mixture Density Networks (\cite{Bishop1994MixtureDN}; see e.g. \cite{Lima_2022} for an application in photo-z estimation).
However NF is already a very general method and overcomes the aforementioned limitations of both NNclass and NNgauss.
% Notice that NF does uses the reparametrization rule to map the target to the base distribution.

\begin{figure*}
    \centering
    \includegraphics[width=1.0\linewidth]{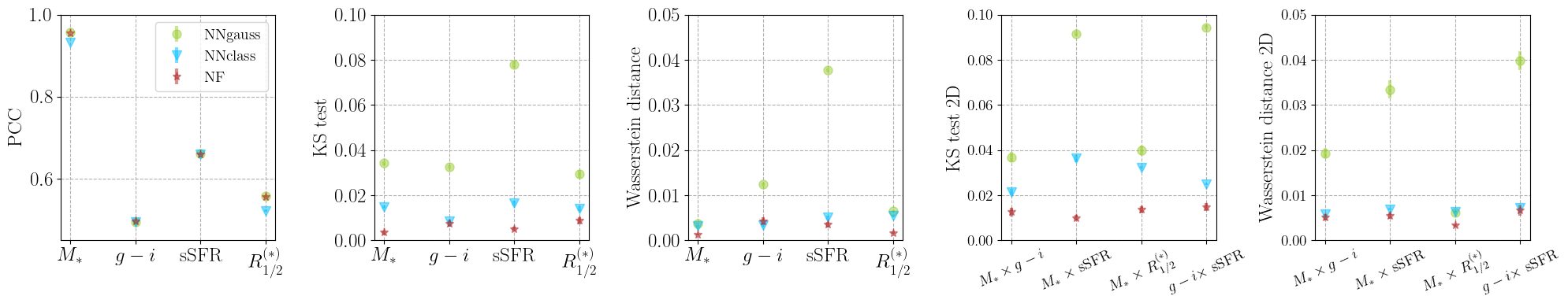}
    \caption{Comparison between the machine learning methods NNgauss, NNclass and NF in terms of the metrics introduced in \S\ref{metrics}. The error bars correspond to the standard deviation over multiple catalogs sampled from the predicted distributions. The 2D scores are computed with a random sub-sample from the test set.}
    \label{fig:comple_sample scores}
\end{figure*}

Fig.\ref{fig:comple_sample scores} shows the metrics introduced in \S\ref{metrics} computed over the generated catalogs. We take the mean and the standard deviation of the scores computed over multiple generated catalogs to obtain the error bars. For the 2D scores we use a randomly selected subset of the test sample (due to computational time).

The PCC is an important metric to verify if the predicted distributions are well-conditioned. Suppose we apply a model that randomly assigns values and simply mimics the scatter and shape of the distribution, i.e., the frequencies of the target values. Even such a model would provide good Wasserstein distance and K-S test on the complete test set. However, since the distributions are not well conditioned, there would be no correlation between true and predicted values, and the PCC score would be poor. On the other hand, if we aim to model the uncertainty in such a way that the sampled catalogs faithfully resemble the reference, it is important to evaluate if we are properly modeling the underlying statistical process that generated the reference catalog, which can be done through the K-S test and the Wasserstein distance.
%On the other hand, a complete analysis is only given in terms of looking at how the methods recover the complete distribution, which is given by the KS-test and Wasserstein distance. 
%This guarantees that a catalog generated with this model will faithfully resemble the reference one. 

We see that, although NNclass shows good performance in terms of the K-S test and Wasserstein distance (comparable to NF results), it is slightly worse than the other methods in terms of PCC. This is perhaps related to the natural limitation of the discretization, which is susceptible to the occupancy and resolution trade-off, and affects the conditioning power of the method.
NNgauss, on the other hand, shows great performance in terms of PCC, similar to NF, although it is not the best assumption to fairly reproduce the details of the shape of the distribution, specially for color and sSFR that have bimodality, as confirmed by the lower performance in terms of the K-S test and Wasserstein distance.

\begin{figure*}%[!h]
    \centering
    \includegraphics[width=0.35\linewidth]{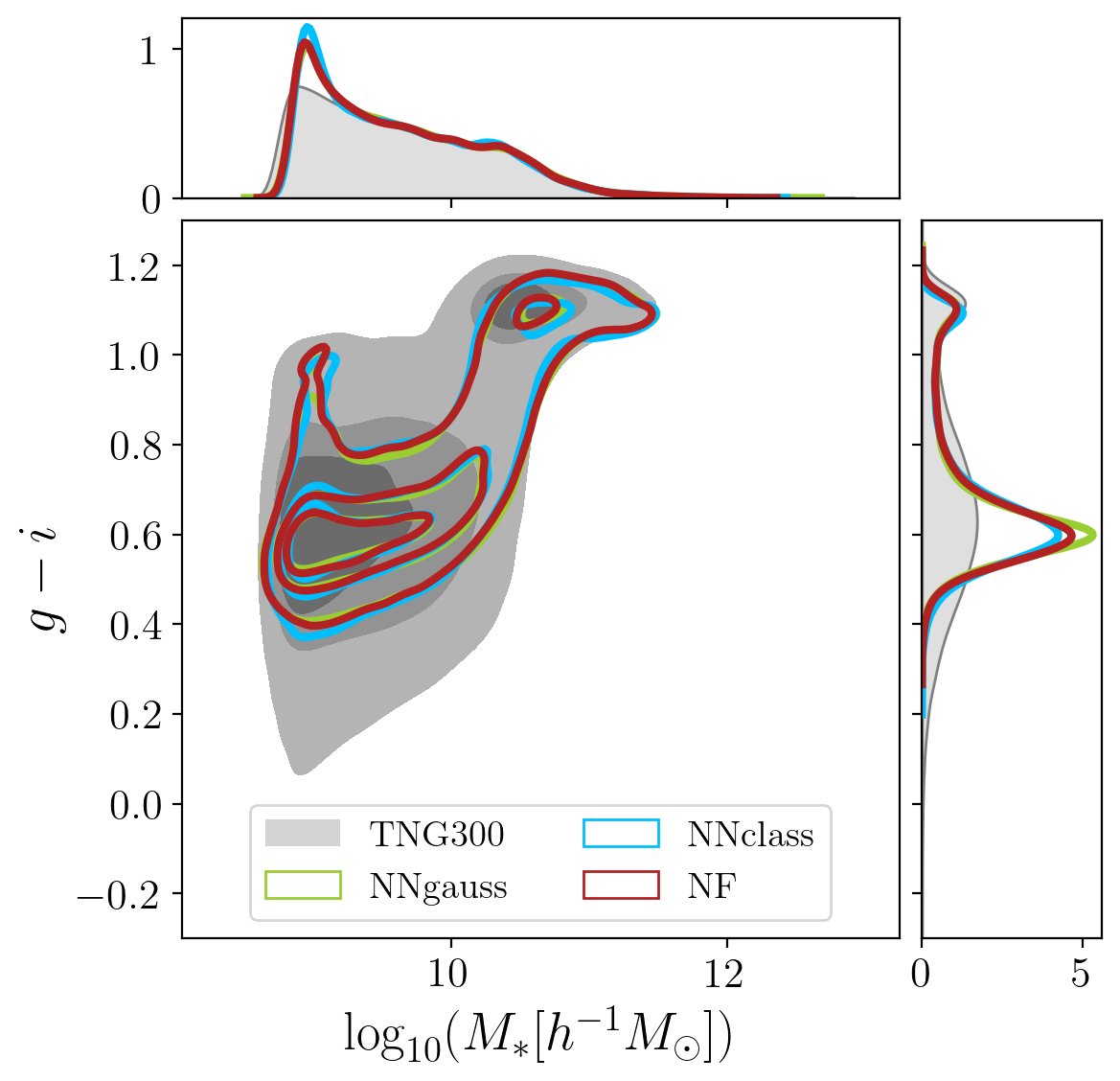}
    \hspace{0.5cm}
    \includegraphics[width=0.35\linewidth]{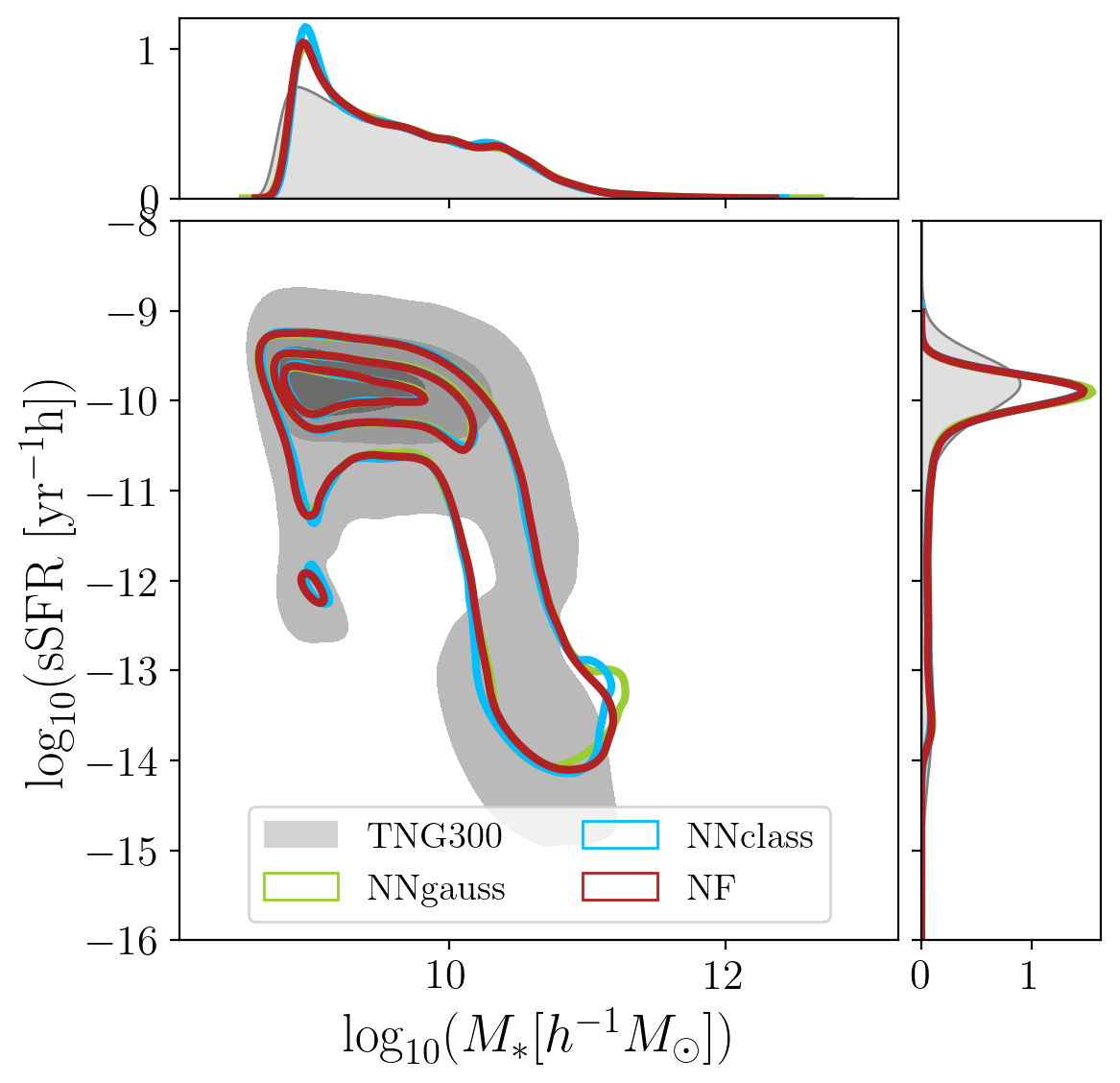} \\
    \includegraphics[width=0.35\linewidth]{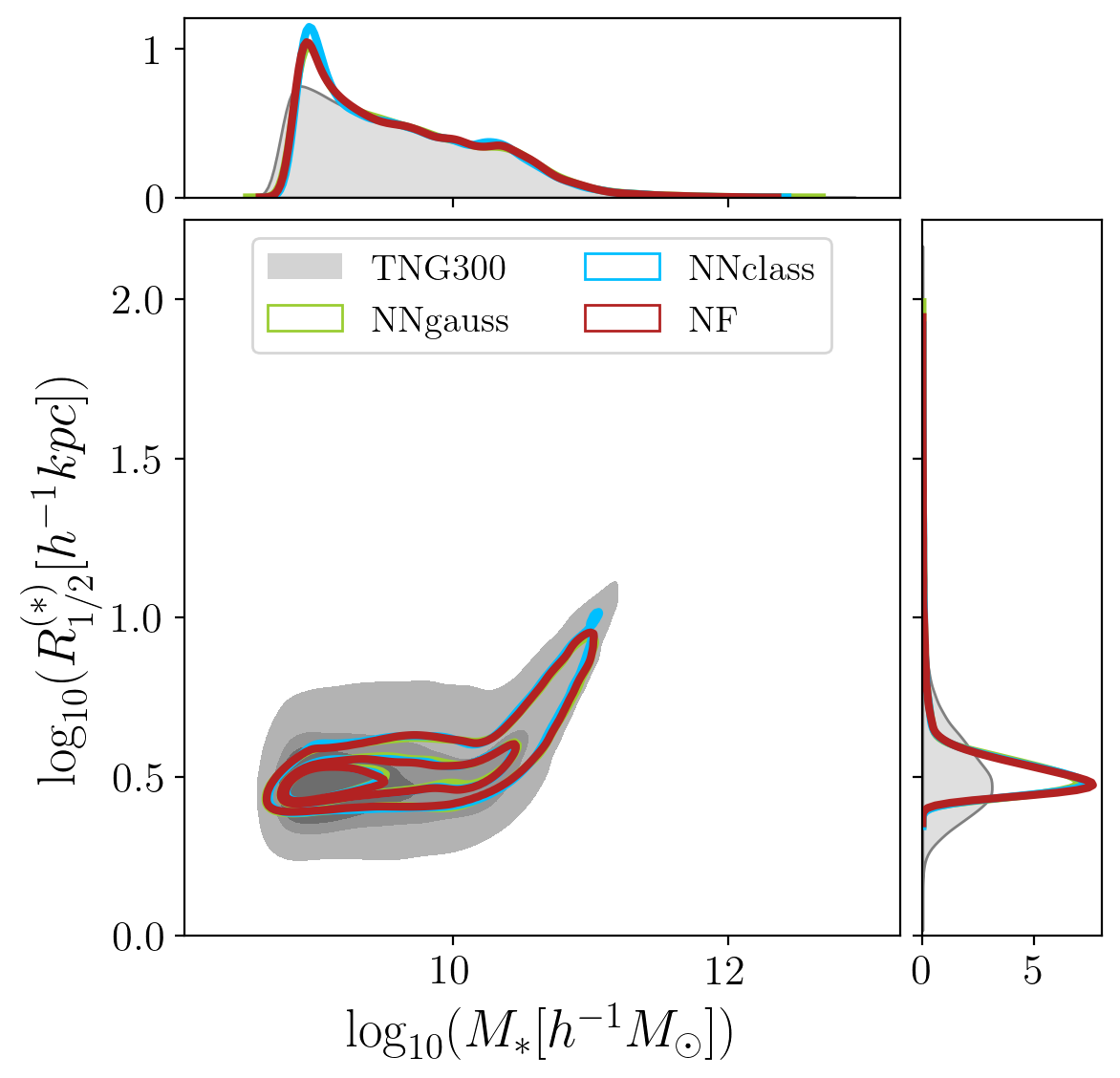}
    \hspace{0.5cm}
    \includegraphics[width=0.35\linewidth]{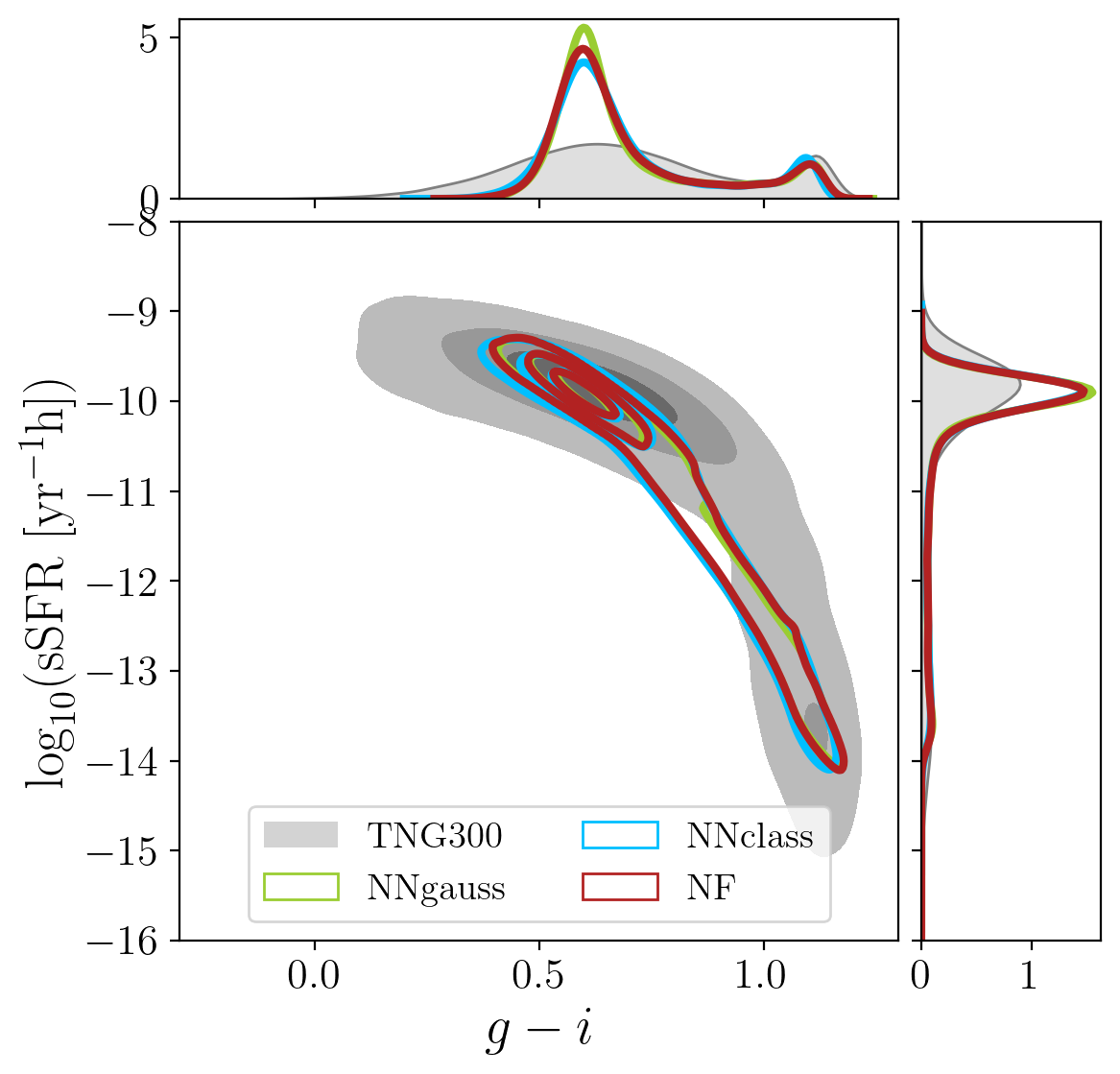}
    \caption{Contour plots of the single-point estimate catalog of the models (NNgauss, NNclass and NF) compared with the truth distribution of the test set from TNG300. This catalog corresponds to the average over 1000 samples of the predicted distribution.}
    \label{fig:maximum likelihood complete_sample scatter}
\end{figure*}

One can also build a single-point estimate catalog based on the expected values by taking the average over multiple samples of the distributions. This should be equivalent to a traditional, deterministic, machine learning estimator like those applied in \cite{deSanti2022}. Fig.\ref{fig:maximum likelihood complete_sample scatter} shows the contour plots of the average over 
%multiple 
1000 samples. As expected, all methods repeat the previous achievements (Raw and SMOGN) resulting in predictions along the peak of the distributions. 
%Therefore, the probabilistic algorithms are consistent with good maximum-likelihood estimators at the same time they fairly reproduce the overall galaxy distributions.

\begin{figure}[!h]
    \centering
    \includegraphics[width=0.8\linewidth]{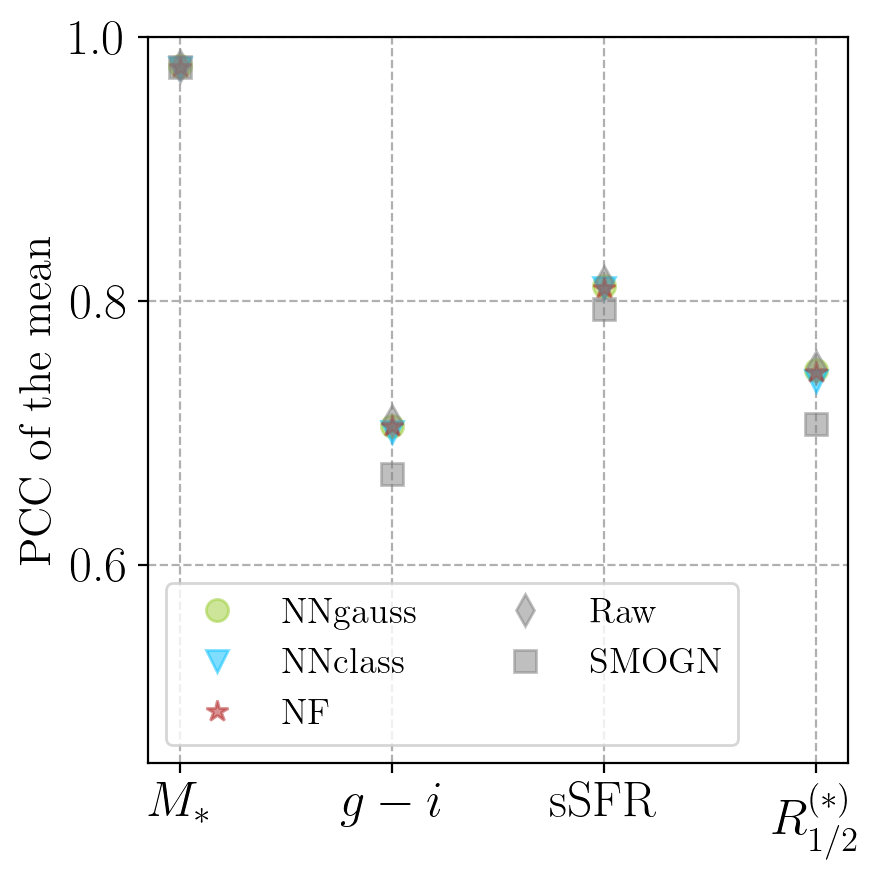}
    \caption{PCC of the single-point estimate catalog %catalogs 
    for each galaxy property. The single-point estimate of NNgauss, NNclass and NF is computed by taking the average over 1000 samples of the predicted distributions.
    }
    %, which are computed by taking the average over 1000 catalogs sampled from the predicted distributions. The Raw and SMOGN methods are deterministic maximum-likelihood estimators.}
    \label{fig:maximum likelihood complete_sample scores}
\end{figure}

Fig. \ref{fig:maximum likelihood complete_sample scores} shows the PCC of the single-point estimate catalogs computed with NNclass, NNgauss, NF, and, for comparison, the deterministic models from \cite{deSanti2022}: Raw (corresponding to the predictions of the stacked models NN, kNN, LGBM, and ERT taken from the raw dataset) and SMOGN (related to the predictions of the stacked models NN, kNN, LGBM, and ERT computed with the SMOGN augmented dataset). 
A similar exercise is done in Appendix A from \cite{Rodrigues2023}.
Due to the intrinsic scatter, we do not expect a high PCC with a single realization of our predicted distributions, as shown in Fig.\ref{fig:comple_sample scores}. 
However, if we take the mean over the 1000 realizations and thus estimate the expected value, the PCC score should be comparable with the deterministic methods.

We see that with all methods we can recover results as good as the Raw model, which demonstrates the flexibility of the probabilistic approach in recovering what is achieved with deterministic estimators. 
We omit the other metrics in this single-point estimation analysis since visual inspection of Fig.\ref{fig:maximum likelihood complete_sample scatter} already shows that the scatter is not well reproduced in this case.

\subsection{Halo populations}\label{halo populations}

In the previous section we discussed the performance of the models in the complete test set. In this section we evaluate the performance on sub-samples of the test set, corresponding to some chosen halo populations, i.e., some selection based on halo properties. 
This analysis is useful to investigate in more detail the conditioning power of the methods, i.e., if they provide the correct populations of galaxies given a population of halos, and to see how well the distributions of different tracers are recovered. They may be more or less degenerate, and therefore show different shapes and scatters, depending on how tightly they are constrained to the halo properties.

To select halo populations, we use HiVAl domains defined over halo parameter space.
As an illustrative case, we do the selection based on the halo mass and age, simply because for 2D distributions we can visualize the HiVAl diagram, but a similar analysis can be done based on a larger number of halo properties
\footnote{We reinforce that to train NNclass we have only used HiVAl in the galaxy parameter space. Not to be confused with the application of HiVAl on the distribution of halo properties.}.

\begin{figure*}%[!h]
    \centering
    \includegraphics[width=0.35\linewidth]{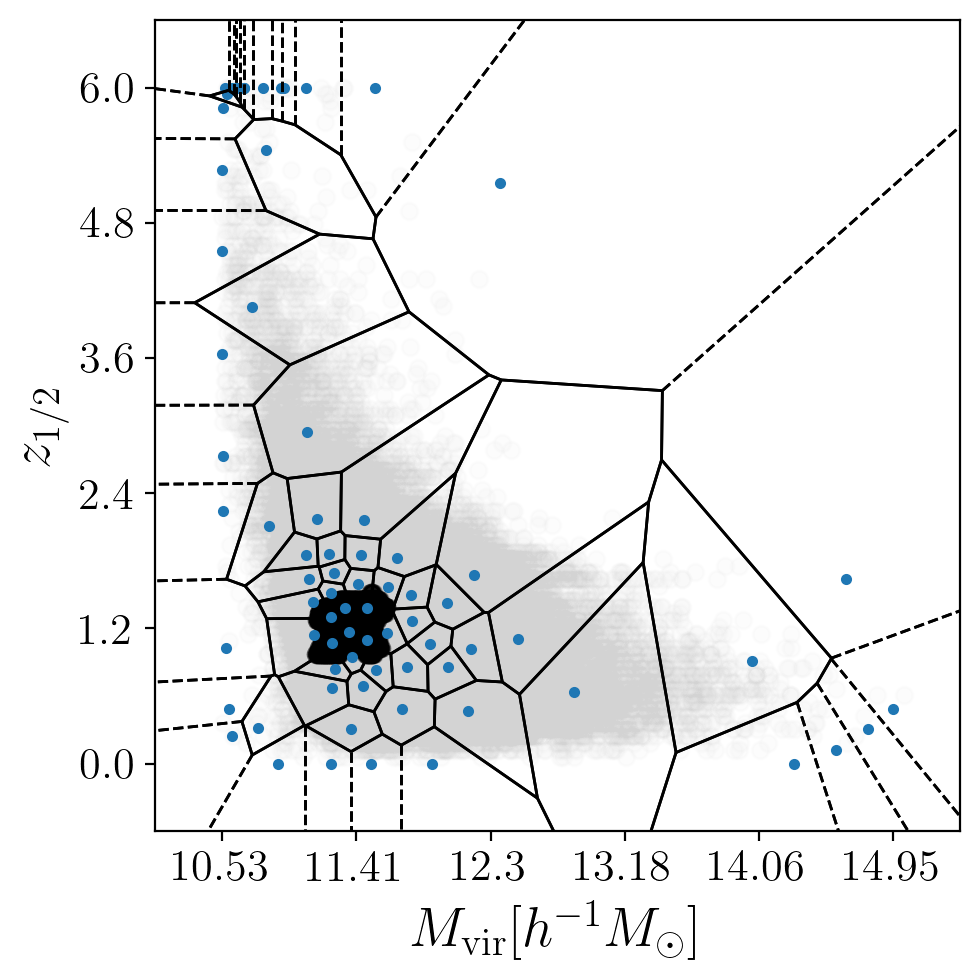}
    \includegraphics[width=0.35\linewidth]{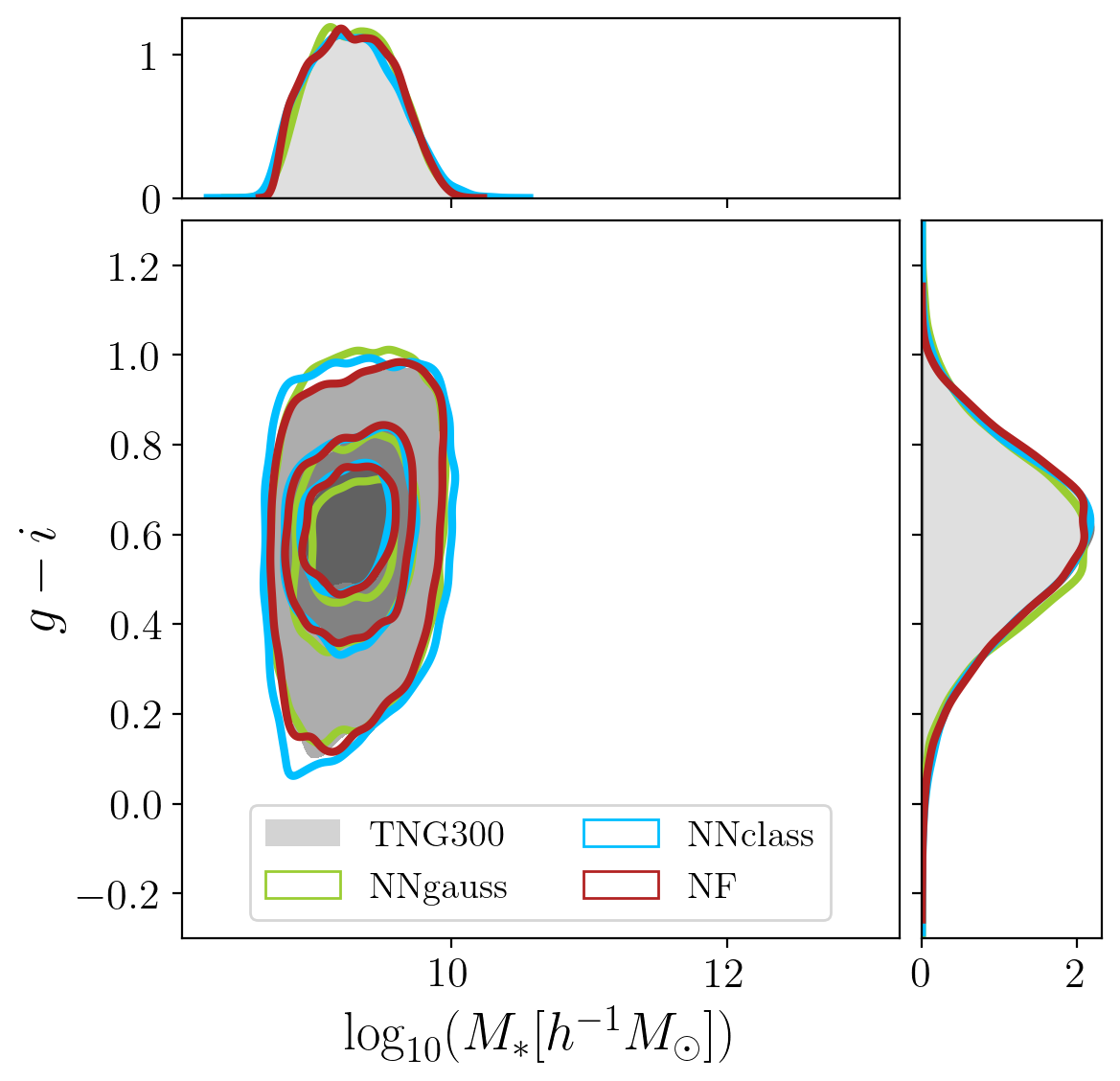}\\
    \includegraphics[width=0.35\linewidth]{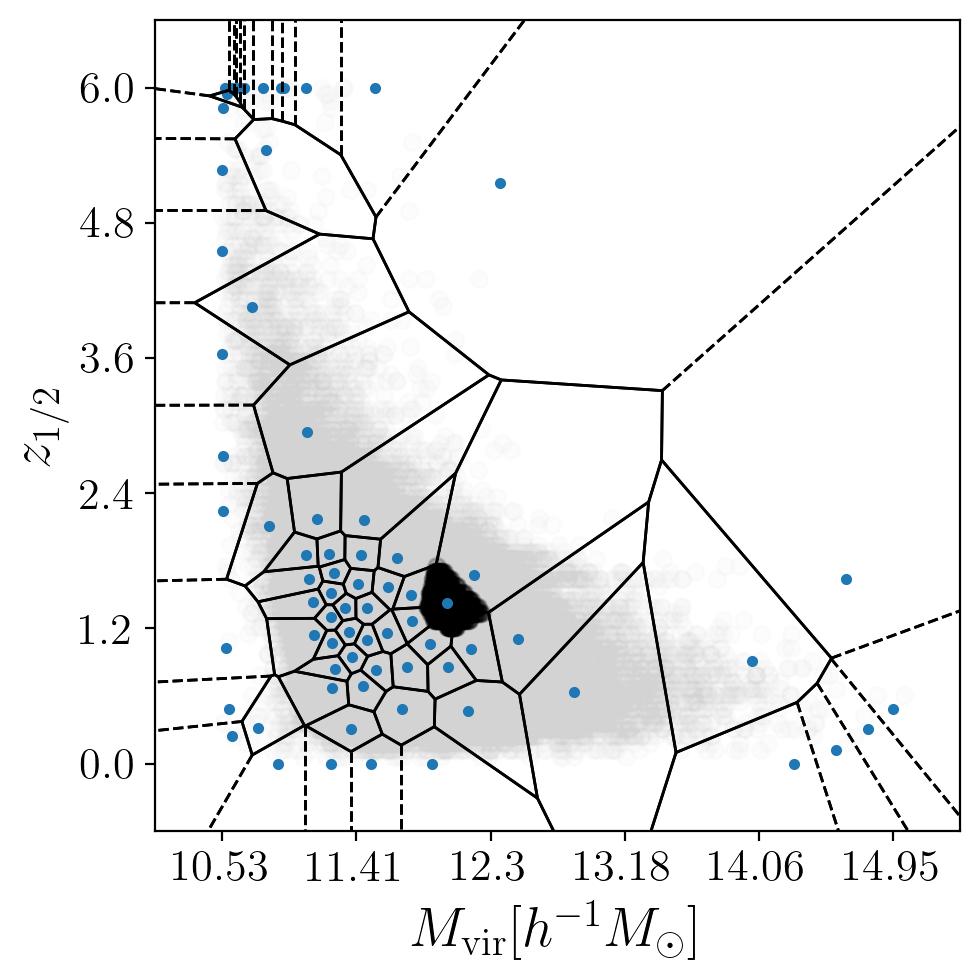}
    \includegraphics[width=0.35\linewidth]{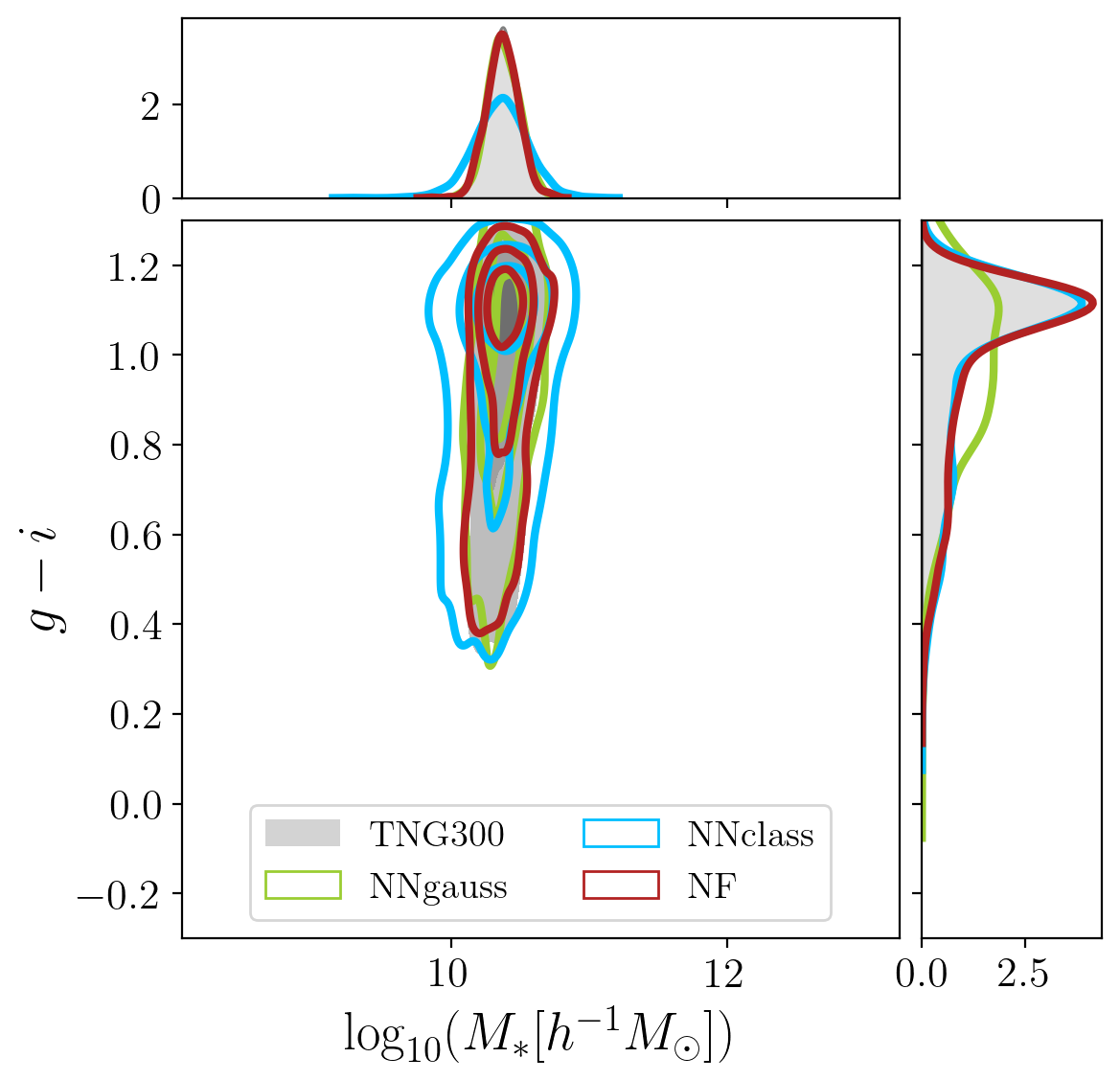}
    \caption{\textit{Left}: two population of halos selected based on HiVAl using halo mass and age. \textit{Right}: color-mass diagram of the galaxies hosted by the corresponding halo populations obtained with TNG300 (reference, gray), NNgauss (green), NNclass (blue) and NF (red).}
    \label{fig:halo populations scatter plots}
\end{figure*}

In the top panel from Fig.\ref{fig:halo populations scatter plots}, we have selected HiVAl domains containing low-mass, intermediate age, halos (total of 10265 instances). The top, left-hand panel shows the HiVAl diagram build upon halo mass and age and highlights the domains containing the selected halo population. The top, right-hand panel shows how the color - stellar mass diagram of the galaxies population the selected halos. Once again, we emphasize that other combinations of properties could be computed, but for the purposes of illustration we keep only the color-mass diagram. Gray regions correspond to the reference TNG300 sample and colored contours correspond to the machine learning models predictions. These halos typically host low-massive, blue galaxies. We see that all models accurately reproduce the properties of the galaxies from these halos. 

The bottom panels from Fig.\ref{fig:halo populations scatter plots} show a similar analysis, but with a different population (total of 1674 instances). As shown in the HiVAl diagram on the bottom, left-hand panel, these halos have intermediate mass and are typically older. This intermediate halo mass regime presents a lot of degeneracy in color, which is likely related to quenching. We see that both blue and red galaxies can occupy those halos. Due to their flexibility, NNclass and NF can accurately predict the range of galaxy colors, while NNgauss does not capture this bimodal feature. Stellar mass, on the other hand, is tightly constrained in this selection and presents a very narrow distribution. Both NF and NNgauss are in good agreement with TNG300 for stellar mass, but NNclass shows strong underfitting.

% \subsection{Single halos}

\section{Conclusions}

As a follow-up of previous works \citep{deSanti2022, Rodrigues2023}, we explore machine learning methods to model the relation between halo and central galaxy properties based on the Illustris TNG300 simulation.

In this work, we investigate methods that model the uncertainty due to the intrinsic stochastic nature of the halo-galaxy connection, and from which one can generate galaxy catalogs that faithfully resemble the reference (TNG300) based on some set of halo properties. This idea has been developed as a proof of concept in \cite{Rodrigues2023} and in the present paper we apply more robust methods that overcome the main limitations of that previous work.

We compare different approaches to model the joint distribution of galaxy properties: a neural network combined with a multivariate Gaussian (NNgauss), a neural network classifier combined with a Categorical distribution (NNclass), and normalizing flows (NF). In particular, we introduce the HiVAl method to define statistically significant regions in a given parameter space based on the density of instances via Voronoi tesselation, which is used to model the target distribution of NNclass.

Each method presents weaknesses and strengths manifested in terms of the different evaluated metrics. We investigate their quality in three main aspects. First, the calibration of the predicted distribution, quantified by the TARP coverage test and by the simulation based calibration rank statistics (see Appendix \ref{add. results}); second, the predictive power of the models, quantified in terms of PCC and through an analysis of single-point estimation; and third, the level of fidelity of the predicted distribution with respect to the reference, quantified in terms of the K-S test and the Wasserstein distance. A summary of the values of the scores is presented in the Appendix \ref{add. results}.

NNgauss shows good predictive power, with results well correlated with the reference - as compared to previous works - but it does not reproduce all the features of the distributions, in particular because they present multimodality. 
NNclass is our first step towards a method able to model a distribution with flexible shape that deviates from Gaussianity. However, it is not optimal due to the discretization of continuous variables, which compromises the predictive power of the method and lead to an underfitting distribution for stellar mass - which should be the best constrained property in halo-galaxy connection, due to its well known high correlation with halo mass. Nevertheless, the satisfactory results in terms of K-S test and the Wasserstein distance suggests that flexible, non-parametric distributions are a promising approach, and motivates the use of NF.
NF accomplishes satisfactory results on all the metrics here evaluated, proving its quality to model the intrinsic statistical process that generated the reference TNG300, and in terms of predictive power.
Moreover, the single-point estimation analysis shows that the models are consistent with the deterministic estimators. 
Therefore, this methodology is a promising auxiliary tool to drive studies and analysis in the context of the halo-galaxy connection, and can serve as a complementary approach to existing methods, such as sub-halo abundance matching (SHAM; see e.g. \citealt{Favole2021, Ortega_Martinez_2024}) and semi-analytical models (SAM; see e.g. \citealt{White1991, Guo2013}).

In order to improve the predictions, one could consider exploring different sets of halo properties. This will be particularly relevant for satellite galaxies, as their formation and evolution involve many complex processes.
Future work will be devoted to explore informative features to predict satellite galaxy properties, combined with the probabilistic approach here proposed. For example, recent works have applied deep learning to optimize feature extraction and thus uncover deterministic correlations hidden in the scatter (see e.g. \cite{Chuang2024}, \cite{2024arXiv240207995W}).

One natural application of this machinery is to populate dark-matter only simulations with TNG300-like galaxies, which reduces the computational cost of running full hydrodynamical simulations.
In \cite{Rodrigues2023} we have demonstrated that the probabilistic model better populates the correct halos with the correct galaxies through an analysis of clustering statistics (specifically with the power spectrum). 
We have quantified how the uncertainty in our halo-galaxy connection model impacts the power spectrum estimation.
We expect that with the methods here presented we will better quantify this error and investigate possible implications for large-scale structure studies.

Another interesting and related application is regarding the inverse problem, i.e., to infer halo properties from galaxy properties, which have a more straightforward application to real observations. 
The exact same machinery could be re-trained to input galaxy properties and output halo properties, i.e., to have a direct modeling of $p({\rm halo}|{\rm gal.})$.

Our training and validation sets are all based on TNG300. In order to apply this to real observations, it is important to stress that we are naturally constrained to whatever limitations that the simulation may have, and some of the stochasticity that we are modeling comes from numerical artifacts of the simulations \citep{Genel_2019}.
Comparisons between simulations with different galaxy formation models should be helpful and informative to evaluate (and possibly extend) the range of applicability of our models (e.g. different sub-grid physics as explored in CAMELS \citep{CAMELS_presentation}).

Finally, we conclude emphasizing that we have selected a few examples to showcase some of the possible analysis that can be done with this machinery and to drive some discussions in the context of halo-galaxy connection.

The pipeline developed in this work is flexible to predict properties individually and jointly, and can be applied in a variety of contexts. 

The material presented in this paper is available in the following \textsc{Github} repository: \href{https://github.com/nvillanova/halo_galaxy_connection_probML}{\url{https://github.com/nvillanova/halo_galaxy_connection_probML}}

\begin{acknowledgements}
NVNR acknowledges financial support from CAPES and the IAC for hospitality during the period which part of this work has been done, financed by PrInt-PDSE grant.
NSMS acknowledges financial support from FAPESP, grants
\href{https://bv.fapesp.br/en/bolsas/187647/cosmological-covariance-matrices-and-machine-learning-methods/}{2019/13108-0} 
and \href{https://bv.fapesp.br/en/bolsas/202438/machine-learning-methods-for-extracting-cosmological-information/}{2022/03589-4}.
RA is supported by FAPESP and CNPq. ADMD thanks Fondecyt for financial support through the Fondecyt Regular 2021 grant 1210612. 
The training, validations and test of the NFs has been carried out using graphic processing units from Simons Foundation, Flatiron Institute, Center of Computational Astrophysics.

\end{acknowledgements}


\bibliographystyle{aa} % style aa.bst
\bibliography{aanda} % your references Yourfile.bib

\begin{thebibliography}{69}
\expandafter\ifx\csname natexlab\endcsname\relax\def\natexlab#1{#1}\fi

\bibitem[{Abadi {et~al.}(2015)Abadi, Agarwal, Barham, Brevdo, Chen, Citro,
  Corrado, Davis, Dean, Devin, Ghemawat, Goodfellow, Harp, Irving, Isard, Jia,
  Jozefowicz, Kaiser, Kudlur, Levenberg, Man\'{e}, Monga, Moore, Murray, Olah,
  Schuster, Shlens, Steiner, Sutskever, Talwar, Tucker, Vanhoucke, Vasudevan,
  Vi\'{e}gas, Vinyals, Warden, Wattenberg, Wicke, Yu, \&
  Zheng}]{tensorflow2015-whitepaper}
Abadi, M., Agarwal, A., Barham, P., {et~al.} 2015, {TensorFlow}: Large-Scale
  Machine Learning on Heterogeneous Systems, software available from
  tensorflow.org

\bibitem[{{Akiba} {et~al.}(2019){Akiba}, {Sano}, {Yanase}, {Ohta}, \&
  {Koyama}}]{optuna2019}
{Akiba}, T., {Sano}, S., {Yanase}, T., {Ohta}, T., \& {Koyama}, M. 2019, arXiv
  e-prints, arXiv:1907.10902

\bibitem[{Arjovsky {et~al.}(2017)Arjovsky, Chintala, \&
  Bottou}]{arjovsky2017wassersteingan}
Arjovsky, M., Chintala, S., \& Bottou, L. 2017, Wasserstein GAN

\bibitem[{{Artale} {et~al.}(2018){Artale}, {Zehavi}, {Contreras}, \&
  {Norberg}}]{Artale2018}
{Artale}, M.~C., {Zehavi}, I., {Contreras}, S., \& {Norberg}, P. 2018, \mnras,
  480, 3978

\bibitem[{Bergstra {et~al.}(2011)Bergstra, Bardenet, Bengio, \&
  K\'{e}gl}]{Bergstra2011}
Bergstra, J., Bardenet, R., Bengio, Y., \& K\'{e}gl, B. 2011, in Advances in
  Neural Information Processing Systems, ed. J.~Shawe-Taylor, R.~Zemel,
  P.~Bartlett, F.~Pereira, \& K.~Weinberger, Vol.~24 (Curran Associates, Inc.)

\bibitem[{Bingham {et~al.}(2019)Bingham, Chen, Jankowiak, Obermeyer, Pradhan,
  Karaletsos, Singh, Szerlip, Horsfall, \& Goodman}]{bingham2019pyro}
Bingham, E., Chen, J.~P., Jankowiak, M., {et~al.} 2019, J. Mach. Learn. Res.,
  20, 28:1

\bibitem[{Bishop(1994)}]{Bishop1994MixtureDN}
Bishop, C. 1994, Mixture density networks, Workingpaper, Aston University

\bibitem[{{Bose} {et~al.}(2019){Bose}, {Eisenstein}, {Hernquist}, {Pillepich},
  {Nelson}, {Marinacci}, {Springel}, \& {Vogelsberger}}]{Bose2019}
{Bose}, S., {Eisenstein}, D.~J., {Hernquist}, L., {et~al.} 2019, \mnras, 2192

\bibitem[{Branco {et~al.}(2017)Branco, Torgo, \& Ribeiro}]{pmlr-v74-branco17a}
Branco, P., Torgo, L., \& Ribeiro, R.~P. 2017, in Proceedings of Machine
  Learning Research, Vol.~74, Proceedings of the First International Workshop
  on Learning with Imbalanced Domains: Theory and Applications, ed. L.~Torgo,
  B.~Krawczyk, P.~Branco, \& N.~Moniz (ECML-PKDD, Skopje, Macedonia: PMLR),
  36--50

\bibitem[{{Bullock} {et~al.}(2001){Bullock}, {Dekel}, {Kolatt}, {Kravtsov},
  {Klypin}, {Porciani}, \& {Primack}}]{Bullock2001_2}
{Bullock}, J.~S., {Dekel}, A., {Kolatt}, T.~S., {et~al.} 2001, \apj, 555, 240

\bibitem[{{Buser}(1978)}]{Buser1978}
{Buser}, R. 1978, \aap, 62, 411

\bibitem[{{Chuang} {et~al.}(2024){Chuang}, {Jespersen}, {Lin}, {Ho}, \&
  {Genel}}]{Chuang2024}
{Chuang}, C.-Y., {Jespersen}, C.~K., {Lin}, Y.-T., {Ho}, S., \& {Genel}, S.
  2024, \apj, 965, 101

\bibitem[{{Coccaro} {et~al.}(2023){Coccaro}, {Letizia}, {Reyes-Gonzalez}, \&
  {Torre}}]{Coccaro2023}
{Coccaro}, A., {Letizia}, M., {Reyes-Gonzalez}, H., \& {Torre}, R. 2023, arXiv
  e-prints, arXiv:2302.12024

\bibitem[{{Contreras} {et~al.}(2020){Contreras}, {Angulo}, \&
  {Zennaro}}]{Contreras2020}
{Contreras}, S., {Angulo}, R., \& {Zennaro}, M. 2020, arXiv e-prints,
  arXiv:2005.03672

\bibitem[{de~Santi {et~al.}(2022)de~Santi, Rodrigues, Montero-Dorta, Abramo,
  Tucci, \& Artale}]{deSanti2022}
de~Santi, N. S.~M., Rodrigues, N. V.~N., Montero-Dorta, A.~D., {et~al.} 2022,
  Monthly Notices of the Royal Astronomical Society, 514, 2463

\bibitem[{{Dillon} {et~al.}(2017){Dillon}, {Langmore}, {Tran}, {Brevdo},
  {Vasudevan}, {Moore}, {Patton}, {Alemi}, {Hoffman}, \&
  {Saurous}}]{2017tensorflowprob}
{Dillon}, J.~V., {Langmore}, I., {Tran}, D., {et~al.} 2017, arXiv e-prints,
  arXiv:1711.10604

\bibitem[{{Dinh} {et~al.}(2014){Dinh}, {Krueger}, \& {Bengio}}]{NICE2014}
{Dinh}, L., {Krueger}, D., \& {Bengio}, Y. 2014, arXiv e-prints,
  arXiv:1410.8516

\bibitem[{{Dinh} {et~al.}(2016){Dinh}, {Sohl-Dickstein}, \&
  {Bengio}}]{RealNVP2016}
{Dinh}, L., {Sohl-Dickstein}, J., \& {Bengio}, S. 2016, arXiv e-prints,
  arXiv:1605.08803

\bibitem[{{Dolatabadi} {et~al.}(2020){Dolatabadi}, {Erfani}, \&
  {Leckie}}]{Dolatabadi2020}
{Dolatabadi}, H.~M., {Erfani}, S., \& {Leckie}, C. 2020, arXiv e-prints,
  arXiv:2001.05168

\bibitem[{{Durkan} {et~al.}(2019){Durkan}, {Bekasov}, {Murray}, \&
  {Papamakarios}}]{NeuroSplines2019}
{Durkan}, C., {Bekasov}, A., {Murray}, I., \& {Papamakarios}, G. 2019, arXiv
  e-prints, arXiv:1906.04032

\bibitem[{{Fasano} \& {Franceschini}(1987)}]{2DKS-Fasano1987}
{Fasano}, G. \& {Franceschini}, A. 1987, \mnras, 225, 155

\bibitem[{{Favole} {et~al.}(2021){Favole}, {Montero-Dorta}, {Artale},
  {Contreras}, {Zehavi}, \& {Xu}}]{Favole2021}
{Favole}, G., {Montero-Dorta}, A.~D., {Artale}, M.~C., {et~al.} 2021, arXiv
  e-prints, arXiv:2101.10733

\bibitem[{Flamary {et~al.}(2021)Flamary, Courty, Gramfort, Alaya, Boisbunon,
  Chambon, Chapel, Corenflos, Fatras, Fournier, Gautheron, Gayraud, Janati,
  Rakotomamonjy, Redko, Rolet, Schutz, Seguy, Sutherland, Tavenard, Tong, \&
  Vayer}]{flamary2021pot}
Flamary, R., Courty, N., Gramfort, A., {et~al.} 2021, Journal of Machine
  Learning Research, 22, 1

\bibitem[{Genel {et~al.}(2019)Genel, Bryan, Springel, Hernquist, Nelson,
  Pillepich, Weinberger, Pakmor, Marinacci, \& Vogelsberger}]{Genel_2019}
Genel, S., Bryan, G.~L., Springel, V., {et~al.} 2019, The Astrophysical
  Journal, 871, 21

\bibitem[{{Genel} {et~al.}(2014){Genel}, {Vogelsberger}, {Springel}, {Sijacki},
  {Nelson}, {Snyder}, {Rodriguez-Gomez}, {Torrey}, \& {Hernquist}}]{Genel2014}
{Genel}, S., {Vogelsberger}, M., {Springel}, V., {et~al.} 2014, \mnras, 445,
  175

\bibitem[{{Gu} {et~al.}(2020){Gu}, {Conroy}, {Diemer}, {Hernquist},
  {Marinacci}, {Nelson}, {Pakmor}, {Pillepich}, \& {Vogelsberger}}]{Gu2020}
{Gu}, M., {Conroy}, C., {Diemer}, B., {et~al.} 2020, arXiv e-prints,
  arXiv:2010.04166

\bibitem[{Guo {et~al.}(2013)Guo, White, Angulo, Henriques, Lemson,
  Boylan-Kolchin, Thomas, \& Short}]{Guo2013}
Guo, Q., White, S., Angulo, R., {et~al.} 2013, Monthly Notices of the Royal
  Astronomical Society, 428, 1351

\bibitem[{{Hadzhiyska} {et~al.}(2021){Hadzhiyska}, {Bose}, {Eisenstein}, \&
  {Hernquist}}]{Hadzhiyska2021}
{Hadzhiyska}, B., {Bose}, S., {Eisenstein}, D., \& {Hernquist}, L. 2021,
  \mnras, 501, 1603

\bibitem[{{Hadzhiyska} {et~al.}(2020){Hadzhiyska}, {Bose}, {Eisenstein},
  {Hernquist}, \& {Spergel}}]{Hadzhiyska2020}
{Hadzhiyska}, B., {Bose}, S., {Eisenstein}, D., {Hernquist}, L., \& {Spergel},
  D.~N. 2020, \mnras, 493, 5506

\bibitem[{Ivezi{\'c} {et~al.}(2014)Ivezi{\'c}, Connolly, VanderPlas, \&
  Gray}]{ivezic2014statistics}
Ivezi{\'c}, {\v{Z}}., Connolly, A., VanderPlas, J., \& Gray, A. 2014,
  Statistics, Data Mining, and Machine Learning in Astronomy: A Practical
  Python Guide for the Analysis of Survey Data, Princeton Series in Modern
  Observational Astronomy (Princeton University Press)

\bibitem[{{Jespersen} {et~al.}(2022){Jespersen}, {Cranmer}, {Melchior}, {Ho},
  {Somerville}, \& {Gabrielpillai}}]{Jespersen2022}
{Jespersen}, C.~K., {Cranmer}, M., {Melchior}, P., {et~al.} 2022, \apj, 941, 7

\bibitem[{{Jo} \& {Kim}(2019)}]{Jo2019}
{Jo}, Y. \& {Kim}, J.-h. 2019, \mnras, 489, 3565

\bibitem[{{Kingma} \& {Ba}(2014)}]{Adam}
{Kingma}, D.~P. \& {Ba}, J. 2014, arXiv e-prints, arXiv:1412.6980

\bibitem[{{Kingma} {et~al.}(2016){Kingma}, {Salimans}, {Jozefowicz}, {Chen},
  {Sutskever}, \& {Welling}}]{auto_regressive-2016}
{Kingma}, D.~P., {Salimans}, T., {Jozefowicz}, R., {et~al.} 2016, arXiv
  e-prints, arXiv:1606.04934

\bibitem[{Kunz(2019)}]{Kunz2019}
Kunz, N. 2019, SMOGN, \url{https://github.com/nickkunz/smogn}

\bibitem[{{Lemos} {et~al.}(2023){Lemos}, {Coogan}, {Hezaveh}, \&
  {Perreault-Levasseur}}]{lemos2023}
{Lemos}, P., {Coogan}, A., {Hezaveh}, Y., \& {Perreault-Levasseur}, L. 2023,
  arXiv e-prints, arXiv:2302.03026

\bibitem[{Lima {et~al.}(2022)Lima, Sodr{\'{e}}, Bom, Teixeira, Nakazono, Buzzo,
  Queiroz, Herpich, Castellon, Dantas, Dors, de~Souza, Akras,
  Jim{\'{e}}nez-Teja, Kanaan, Ribeiro, \& Schoennell}]{Lima_2022}
Lima, E., Sodr{\'{e}}, L., Bom, C., {et~al.} 2022, Astronomy and Computing, 38,
  100510

\bibitem[{{Lovell} {et~al.}(2023){Lovell}, {Hassan}, {Villaescusa-Navarro},
  {Genel}, {Hahn}, {Angles-Alcazar}, {Kwon}, {de Santi}, {Iyer}, {Fabbian}, \&
  {Bryan}}]{Lovell2023}
{Lovell}, C.~C., {Hassan}, S., {Villaescusa-Navarro}, F., {et~al.} 2023, in
  Machine Learning for Astrophysics, 21

\bibitem[{{Marinacci} {et~al.}(2018){Marinacci}, {Vogelsberger}, {Pakmor},
  {Torrey}, {Springel}, {Hernquist}, {Nelson}, {Weinberger}, {Pillepich},
  {Naiman}, \& {Genel}}]{Marinacci2018}
{Marinacci}, F., {Vogelsberger}, M., {Pakmor}, R., {et~al.} 2018, \mnras, 480,
  5113

\bibitem[{{Montero-Dorta} {et~al.}(2021{\natexlab{a}}){Montero-Dorta},
  {Artale}, {Abramo}, \& {Tucci}}]{MonteroDorta2021A}
{Montero-Dorta}, A.~D., {Artale}, M.~C., {Abramo}, L.~R., \& {Tucci}, B.
  2021{\natexlab{a}}, \mnras, 504, 4568

\bibitem[{{Montero-Dorta} {et~al.}(2020){Montero-Dorta}, {Artale}, {Abramo},
  {Tucci}, {Padilla}, {Sato-Polito}, {Lacerna}, {Rodriguez}, \&
  {Angulo}}]{MonteroDorta2020}
{Montero-Dorta}, A.~D., {Artale}, M.~C., {Abramo}, L.~R., {et~al.} 2020,
  \mnras, 496, 1182

\bibitem[{{Montero-Dorta} {et~al.}(2021{\natexlab{b}}){Montero-Dorta},
  {Chaves-Montero}, {Artale}, \& {Favole}}]{MonteroDorta2021B}
{Montero-Dorta}, A.~D., {Chaves-Montero}, J., {Artale}, M.~C., \& {Favole}, G.
  2021{\natexlab{b}}, \mnras, 508, 940

\bibitem[{{Montero-Dorta} \& {Rodriguez}(2024)}]{MonteroDorta2024}
{Montero-Dorta}, A.~D. \& {Rodriguez}, F. 2024, \mnras, 531, 290

\bibitem[{{Naiman} {et~al.}(2018){Naiman}, {Pillepich}, {Springel},
  {Ramirez-Ruiz}, {Torrey}, {Vogelsberger}, {Pakmor}, {Nelson}, {Marinacci},
  {Hernquist}, {Weinberger}, \& {Genel}}]{Naiman2018}
{Naiman}, J.~P., {Pillepich}, A., {Springel}, V., {et~al.} 2018, \mnras, 477,
  1206

\bibitem[{{Navarro} {et~al.}(1997){Navarro}, {Frenk}, \& {White}}]{nfw1997}
{Navarro}, J.~F., {Frenk}, C.~S., \& {White}, S. D.~M. 1997, \apj, 490, 493

\bibitem[{{Nelson} {et~al.}(2018){Nelson}, {Pillepich}, {Springel},
  {Weinberger}, {Hernquist}, {Pakmor}, {Genel}, {Torrey}, {Vogelsberger},
  {Kauffmann}, {Marinacci}, \& {Naiman}}]{Nelson2018_ColorBim}
{Nelson}, D., {Pillepich}, A., {Springel}, V., {et~al.} 2018, \mnras, 475, 624

\bibitem[{Nelson {et~al.}(2019)Nelson, Springel, Pillepich, \&
  et~al.}]{Nelson2019}
Nelson, D., Springel, V., Pillepich, A., \& et~al. 2019, Computational
  Astrophysics and Cosmology, 6

\bibitem[{{Neyrinck}(2008)}]{ZOBOV}
{Neyrinck}, M.~C. 2008, \mnras, 386, 2101

\bibitem[{Ortega-Martinez {et~al.}(2024)Ortega-Martinez, Contreras, \&
  Angulo}]{Ortega_Martinez_2024}
Ortega-Martinez, S., Contreras, S., \& Angulo, R. 2024, Astronomy \&amp;
  Astrophysics, 689, A66

\bibitem[{Papamakarios {et~al.}(2021)Papamakarios, Nalisnick, Rezende, Mohamed,
  \& Lakshminarayanan}]{JMLR:v22:19-1028}
Papamakarios, G., Nalisnick, E., Rezende, D.~J., Mohamed, S., \&
  Lakshminarayanan, B. 2021, Journal of Machine Learning Research, 22, 1

\bibitem[{{Peacock}(1983)}]{2DKS-Peacock1983}
{Peacock}, J.~A. 1983, \mnras, 202, 615

\bibitem[{{Pillepich} {et~al.}(2018{\natexlab{a}}){Pillepich}, {Nelson},
  {Hernquist}, {Springel}, {Pakmor}, {Torrey}, {Weinberger}, {Genel}, {Naiman},
  {Marinacci}, \& {Vogelsberger}}]{Pillepich2018b}
{Pillepich}, A., {Nelson}, D., {Hernquist}, L., {et~al.} 2018{\natexlab{a}},
  \mnras, 475, 648

\bibitem[{{Pillepich} {et~al.}(2018{\natexlab{b}}){Pillepich}, {Springel},
  {Nelson}, {Genel}, {Naiman}, {Pakmor}, {Hernquist}, {Torrey}, {Vogelsberger},
  {Weinberger}, \& {Marinacci}}]{Pillepich2018}
{Pillepich}, A., {Springel}, V., {Nelson}, D., {et~al.} 2018{\natexlab{b}},
  \mnras, 473, 4077

\bibitem[{{Planck Collaboration} {et~al.}(2016){Planck Collaboration}, {Ade},
  {Aghanim}, {Arnaud}, {Ashdown}, {Aumont}, {Baccigalupi}, {Banday},
  {Barreiro}, {Bartlett}, {Bartolo}, {Battaner}, {Battye}, {Benabed},
  {Beno{\^\i}t}, {Benoit-L{\'e}vy}, {Bernard}, {Bersanelli}, {Bielewicz},
  {Bock}, {Bonaldi}, {Bonavera}, {Bond}, {Borrill}, {Bouchet}, {Boulanger},
  {Bucher}, {Burigana}, {Butler}, {Calabrese}, {Cardoso}, {Catalano},
  {Challinor}, {Chamballu}, {Chary}, {Chiang}, {Chluba}, {Christensen},
  {Church}, {Clements}, {Colombi}, {Colombo}, {Combet}, {Coulais}, {Crill},
  {Curto}, {Cuttaia}, {Danese}, {Davies}, {Davis}, {de Bernardis}, {de Rosa},
  {de Zotti}, {Delabrouille}, {D{\'e}sert}, {Di Valentino}, {Dickinson},
  {Diego}, {Dolag}, {Dole}, {Donzelli}, {Dor{\'e}}, {Douspis}, {Ducout},
  {Dunkley}, {Dupac}, {Efstathiou}, {Elsner}, {En{\ss}lin}, {Eriksen},
  {Farhang}, {Fergusson}, {Finelli}, {Forni}, {Frailis}, {Fraisse},
  {Franceschi}, {Frejsel}, {Galeotta}, {Galli}, {Ganga}, {Gauthier}, {Gerbino},
  {Ghosh}, {Giard}, {Giraud-H{\'e}raud}, {Giusarma}, {Gjerl{\o}w},
  {Gonz{\'a}lez-Nuevo}, {G{\'o}rski}, {Gratton}, {Gregorio}, {Gruppuso},
  {Gudmundsson}, {Hamann}, {Hansen}, {Hanson}, {Harrison}, {Helou},
  {Henrot-Versill{\'e}}, {Hern{\'a}ndez-Monteagudo}, {Herranz}, {Hildebrandt},
  {Hivon}, {Hobson}, {Holmes}, {Hornstrup}, {Hovest}, {Huang}, {Huffenberger},
  {Hurier}, {Jaffe}, {Jaffe}, {Jones}, {Juvela}, {Keih{\"a}nen}, {Keskitalo},
  {Kisner}, {Kneissl}, {Knoche}, {Knox}, {Kunz}, {Kurki-Suonio}, {Lagache},
  {L{\"a}hteenm{\"a}ki}, {Lamarre}, {Lasenby}, {Lattanzi}, {Lawrence}, {Leahy},
  {Leonardi}, {Lesgourgues}, {Levrier}, {Lewis}, {Liguori}, {Lilje},
  {Linden-V{\o}rnle}, {L{\'o}pez-Caniego}, {Lubin}, {Mac{\'\i}as-P{\'e}rez},
  {Maggio}, {Maino}, {Mandolesi}, {Mangilli}, {Marchini}, {Maris}, {Martin},
  {Martinelli}, {Mart{\'\i}nez-Gonz{\'a}lez}, {Masi}, {Matarrese}, {McGehee},
  {Meinhold}, {Melchiorri}, {Melin}, {Mendes}, {Mennella}, {Migliaccio},
  {Millea}, {Mitra}, {Miville-Desch{\^e}nes}, {Moneti}, {Montier}, {Morgante},
  {Mortlock}, {Moss}, {Munshi}, {Murphy}, {Naselsky}, {Nati}, {Natoli},
  {Netterfield}, {N{\o}rgaard-Nielsen}, {Noviello}, {Novikov}, {Novikov},
  {Oxborrow}, {Paci}, {Pagano}, {Pajot}, {Paladini}, {Paoletti}, {Partridge},
  {Pasian}, {Patanchon}, {Pearson}, {Perdereau}, {Perotto}, {Perrotta},
  {Pettorino}, {Piacentini}, {Piat}, {Pierpaoli}, {Pietrobon}, {Plaszczynski},
  {Pointecouteau}, {Polenta}, {Popa}, {Pratt}, {Pr{\'e}zeau}, {Prunet},
  {Puget}, {Rachen}, {Reach}, {Rebolo}, {Reinecke}, {Remazeilles}, {Renault},
  {Renzi}, {Ristorcelli}, {Rocha}, {Rosset}, {Rossetti}, {Roudier},
  {Rouill{\'e} d'Orfeuil}, {Rowan-Robinson}, {Rubi{\~n}o-Mart{\'\i}n},
  {Rusholme}, {Said}, {Salvatelli}, {Salvati}, {Sandri}, {Santos},
  {Savelainen}, {Savini}, {Scott}, {Seiffert}, {Serra}, {Shellard}, {Spencer},
  {Spinelli}, {Stolyarov}, {Stompor}, {Sudiwala}, {Sunyaev}, {Sutton},
  {Suur-Uski}, {Sygnet}, {Tauber}, {Terenzi}, {Toffolatti}, {Tomasi},
  {Tristram}, {Trombetti}, {Tucci}, {Tuovinen}, {T{\"u}rler}, {Umana},
  {Valenziano}, {Valiviita}, {Van Tent}, {Vielva}, {Villa}, {Wade}, {Wandelt},
  {Wehus}, {White}, {White}, {Wilkinson}, {Yvon}, {Zacchei}, \&
  {Zonca}}]{Planck2016}
{Planck Collaboration}, {Ade}, P.~A.~R., {Aghanim}, N., {et~al.} 2016, \aap,
  594, A13

\bibitem[{{Ramdas} {et~al.}(2015){Ramdas}, {Garcia}, \&
  {Cuturi}}]{Wasserstein2015}
{Ramdas}, A., {Garcia}, N., \& {Cuturi}, M. 2015, arXiv e-prints,
  arXiv:1509.02237

\bibitem[{{Rodrigues} {et~al.}(2023){Rodrigues}, {de Santi}, {Montero-Dorta},
  \& {Abramo}}]{Rodrigues2023}
{Rodrigues}, N. V.~N., {de Santi}, N. S.~M., {Montero-Dorta}, A.~D., \&
  {Abramo}, L.~R. 2023, \mnras, 522, 3236

\bibitem[{{Shi} {et~al.}(2020){Shi}, {Wang}, {Mo}, {Vogelsberger}, {Ho}, {Du},
  {Nelson}, {Pillepich}, \& {Hernquist}}]{Shi2020}
{Shi}, J., {Wang}, H., {Mo}, H., {et~al.} 2020, \apj, 893, 139

\bibitem[{{Springel}(2010)}]{Springel2010}
{Springel}, V. 2010, \mnras, 401, 791

\bibitem[{{Springel} {et~al.}(2018){Springel}, {Pakmor}, {Pillepich},
  {Weinberger}, {Nelson}, {Hernquist}, {Vogelsberger}, {Genel}, {Torrey},
  {Marinacci}, \& {Naiman}}]{Springel2018}
{Springel}, V., {Pakmor}, R., {Pillepich}, A., {et~al.} 2018, \mnras, 475, 676

\bibitem[{{Stiskalek} {et~al.}(2022){Stiskalek}, {Bartlett}, {Desmond}, \&
  {Anbajagane}}]{Stiskalek2022}
{Stiskalek}, R., {Bartlett}, D.~J., {Desmond}, H., \& {Anbajagane}, D. 2022,
  \mnras, 514, 4026

\bibitem[{Taillon(2018)}]{2DKS}
Taillon, G. 2018, 2DKS, \url{https://github.com/Gabinou/2DKS}

\bibitem[{Talts {et~al.}(2020)Talts, Betancourt, Simpson, Vehtari, \&
  Gelman}]{sbc}
Talts, S., Betancourt, M., Simpson, D., Vehtari, A., \& Gelman, A. 2020,
  Validating Bayesian Inference Algorithms with Simulation-Based Calibration

\bibitem[{{Villaescusa-Navarro} {et~al.}(2021){Villaescusa-Navarro},
  {Angl{\'e}s-Alc{\'a}zar}, {Genel}, {Spergel}, {Somerville}, {Dave},
  {Pillepich}, {Hernquist}, {Nelson}, {Torrey}, {Narayanan}, {Li}, {Philcox},
  {La Torre}, {Maria Delgado}, {Ho}, {Hassan}, {Burkhart}, {Wadekar},
  {Battaglia}, {Contardo}, \& {Bryan}}]{CAMELS_presentation}
{Villaescusa-Navarro}, F., {Angl{\'e}s-Alc{\'a}zar}, D., {Genel}, S., {et~al.}
  2021, \apj, 915, 71

\bibitem[{Virtanen {et~al.}(2020)Virtanen, Gommers, Oliphant, Haberland, Reddy,
  Cournapeau, Burovski, Peterson, Weckesser, Bright, {van der Walt}, Brett,
  Wilson, Millman, Mayorov, Nelson, Jones, Kern, Larson, Carey, Polat, Feng,
  Moore, {VanderPlas}, Laxalde, Perktold, Cimrman, Henriksen, Quintero, Harris,
  Archibald, Ribeiro, Pedregosa, {van Mulbregt}, \& {SciPy 1.0
  Contributors}}]{2020SciPy-NMeth}
Virtanen, P., Gommers, R., Oliphant, T.~E., {et~al.} 2020, Nature Methods, 17,
  261

\bibitem[{{Vogelsberger} {et~al.}(2014{\natexlab{a}}){Vogelsberger}, {Genel},
  {Springel}, {Torrey}, {Sijacki}, {Xu}, {Snyder}, {Bird}, {Nelson}, \&
  {Hernquist}}]{Vogelsberger2014b}
{Vogelsberger}, M., {Genel}, S., {Springel}, V., {et~al.} 2014{\natexlab{a}},
  \nat, 509, 177

\bibitem[{{Vogelsberger} {et~al.}(2014{\natexlab{b}}){Vogelsberger}, {Genel},
  {Springel}, {Torrey}, {Sijacki}, {Xu}, {Snyder}, {Nelson}, \&
  {Hernquist}}]{Vogelsberger2014a}
{Vogelsberger}, M., {Genel}, S., {Springel}, V., {et~al.} 2014{\natexlab{b}},
  \mnras, 444, 1518

\bibitem[{{Wechsler} \& {Tinker}(2018)}]{Wechsler2018}
{Wechsler}, R.~H. \& {Tinker}, J.~L. 2018, \araa, 56, 435

\bibitem[{White \& Frenk(1991)}]{White1991}
White, S. D.~M. \& Frenk, C.~S. 1991, Astrophysical Journal, 379, 52

\bibitem[{{Wu} {et~al.}(2024){Wu}, {Kragh Jespersen}, \&
  {Wechsler}}]{2024arXiv240207995W}
{Wu}, J.~F., {Kragh Jespersen}, C., \& {Wechsler}, R.~H. 2024, arXiv e-prints,
  arXiv:2402.07995

\end{thebibliography}

\begin{appendix}

\section{Additional results}\label{add. results}

\subsection{Simulation based calibration}
\begin{figure}%[!h]
    \centering
    \includegraphics[width=0.495\linewidth]{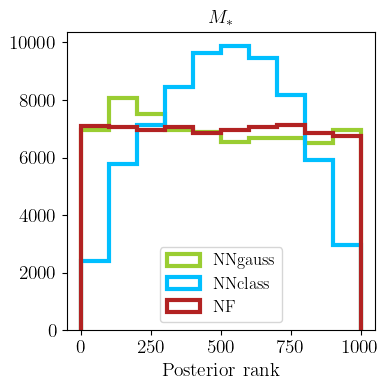}
    % \hspace{0.5cm}
    \includegraphics[width=0.495\linewidth]{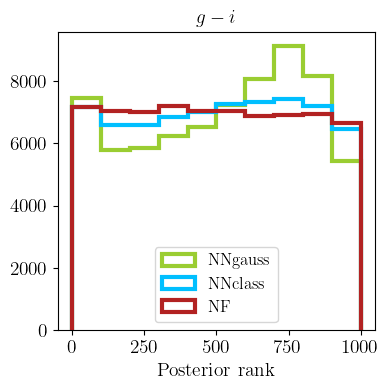}\\
    \includegraphics[width=0.495\linewidth]{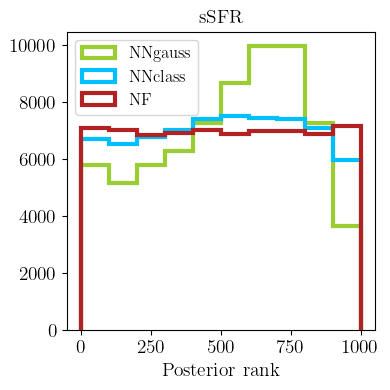}
    % \hspace{0.5cm}
    \includegraphics[width=0.495\linewidth]{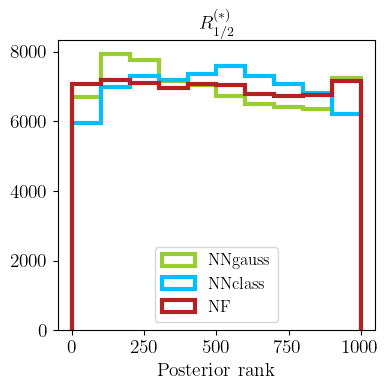}
    \caption{Simulation based calibration rank statistics (see \S\ref{sbc score}) computed for each galaxy property and for each ML model.}
    \label{fig:sbc rank}
\end{figure}

Fig.\ref{fig:sbc rank} shows the SBC rank statistics for each galaxy property and for each ML method. As discussed in \S\ref{sbc score}, a good model should present this score following a Uniform distribution. It is computed with 1000 samples per instance, for all test set instances. 
By visual inspection, we see that NF shows good agreement for all properties, once more validating the quality of the predicted distribution. 
The observed pattern for NNclass' stellar mass prediction suggest that the model is underfitting, which is in agreement with the TARP coverage test result shown in Fig.\ref{fig:comple_sample CT}. 
For NNgauss we see once again more pronounced deviations indicating a biased prediction for color and sSFR, which are the properties that present bimodality. 

\subsection{Tables}

This Section contains the Tables comparing the values of the scores presented in \S\ref{results}. The best values for each score and each galaxy property are highlighted in bold.

Table \ref{table:pcc average} contains the PCC scores shown in Figure \ref{fig:maximum likelihood complete_sample scores}. The best scores for all galaxy properties are found for NNgauss, followed by NF predictions.
Table \ref{table:pcc} contains the PCC scores shown in Figure \ref{fig:comple_sample scores}.
In this case, NNgauss provides the best predictions for stellar mass, sSFR, and radius while NF better predicts galaxy color.

Tables \ref{table:ks1d} and \ref{table:ks2d} contain the values of the 1D and 2D K-S test, respectively (see Fig.\ref{fig:comple_sample scores}).
We have implemented our own 1D K-S test schedule and used the \cite{2DKS} repository to compute the 2D K-S test. 
For the K-S test the best scores follow for NF predictions.

Tables \ref{table:wd1d} and \ref{table:wd2d} contain the values of the 1D and 2D Wasserstein distance, respectively (see Fig.\ref{fig:comple_sample scores}). 
We compute the 1D Wasserstein distance based on \cite{2020SciPy-NMeth} and the 2D version using \cite{flamary2021pot}. 
Once again NF provides the best scores but now for the 2D Wasserstein, while it holds the best values for the 1D version, unless for color predictions, when NNclass produces the best results.

\begin{table*}[h]
\caption{Pearson correlation coefficient between the reference and predicted sample for each galaxy property. The predicted value is the average over 1000 samples per instance in the test set.}             % title of Table
\label{table:pcc average}      % is used to refer this table in the text
\centering                          % used for centering table

% \begin{tabular}{l c c c c}        % centered columns (4 columns)
% \hline\hline                 % inserts double horizontal lines
%  PCC avg & $M_*$ & $g - i$ & sSFR & $R_{1/2}^{(*)}$ \\    % table heading 
% \hline                        % inserts single horizontal line
%     NNgauss & \textbf{0.978} & \textbf{0.706} & \textbf{0.811} & \textbf{0.748} \\
%     NNclass & 0.977 & 0.701 & 0.810 & 0.739 \\ 
%     NF & \textbf{0.978} & \textbf{0.706} & \textbf{0.811} & 0.747 \\
% \hline                                   %inserts single line
% \end{tabular}
% \end{table*}

\begin{tabular}{l c c c c}        % centered columns (4 columns)
\hline\hline                 % inserts double horizontal lines
 PCC avg & $M_*$ & $g - i$ & sSFR & $R_{1/2}^{(*)}$ \\    % table heading 
\hline                        % inserts single horizontal line
    NNgauss & \textbf{0.9781} & 0.7056 & \textbf{0.8112} & \textbf{0.7477} \\
    NNclass & 0.9769 & 0.7010 & 0.8101 & 0.7394 \\ 
    NF & 0.9779 & \textbf{0.7057} & 0.8105 & 0.7468 \\
\hline                                   %inserts single line
\end{tabular}
\end{table*}

\begin{table*}[h]
\caption{Pearson correlation coefficient between the reference and predicted sample for each galaxy property. The mean and standard deviation are computed with 1000 samples per instance in the test set.}             % title of Table
\label{table:pcc}      % is used to refer this table in the text
\centering                          % used for centering table

\begin{tabular}{l c c c c l c c c c}        % centered columns (4 columns)
\hline\hline                 % inserts double horizontal lines
 PCC & $M_*$ & $g - i$ & sSFR & $R_{1/2}^{(*)}$\\    % table heading 
\hline                        % inserts single horizontal line
    NNgauss & $\textbf{0.9578} \pm 0.0002$ & $0.4938 \pm 0.0023$ & $\textbf{0.6611} \pm 0.0023$ & $\textbf{0.5583} \pm 0.0021$ \\
    NNclass & $0.9320 \pm 0.0004$ & $0.4939 \pm 0.0022$ & $0.6591 \pm 0.0027$ & $0.5223 \pm 0.0026$ \\
    NF      & $0.9554 \pm 0.0002$ & $\textbf{0.4971} \pm 0.0023$ & $0.6597 \pm 0.0024$ & $0.5557 \pm 0.0022$ \\
\hline                                   %inserts single line
\end{tabular}
\end{table*}

\begin{table*}
\caption{1D K-S test computed for each galaxy property. 1D scores are computed with the complete test set catalog, and 1000 samples per instance to calculate the standard deviation.}             % title of Table
\label{table:ks1d}      % is used to refer this table in the text
\centering                          % used for centering table

\begin{tabular}{l c c c c}        % centered columns (4 columns)
\hline\hline                 % inserts double horizontal lines
1D K-S test & $M_*$ & $g - i$ & sSFR & $R_{1/2}^{(*)}$\\    % table heading 
\hline                        % inserts single horizontal line
    NNgauss & $0.0342 \pm 0.0009$ & $0.0326 \pm 0.0007$ & $0.0780 \pm 0.0015$ & $0.0294 \pm 0.0017$ \\
    NNclass & $0.0147 \pm 0.0008$ & $0.0085 \pm 0.0005$ & $0.0164 \pm 0.0010$ & $0.0140 \pm 0.0009$ \\
    NF & $\mathbf{0.0034} \pm 0.0008$ & $\mathbf{0.0073} \pm 0.0014$ & $\mathbf{0.0050} \pm 0.0008$ & $\mathbf{0.0086} \pm 0.0016$ \\
\hline                                   %inserts single line
\end{tabular}
\end{table*}

\begin{table*}
\caption{2D K-S test computed for each pair of galaxy properties. 2D scores are computed using a randomly drawn subset of $3\times10^4$ instances from the test set, and 5 samples per instance to compute the standard deviation.}             % title of Table
\label{table:ks2d}      % is used to refer this table in the text
\centering                          % used for centering table

\begin{tabular}{l c c c c}        % centered columns (4 columns)
\hline\hline                 % inserts double horizontal lines
2D K-S test & $M_* \times g - i$ & $M_* \times $sSFR & $M_* \times R_{1/2}^{(*)}$ & $g - i \times $sSFR\\    % table heading 
\hline                        % inserts single horizontal line
    NNgauss & $0.0368 \pm 0.0022$ & $0.0915 \pm 0.0010$ & $0.0399 \pm 0.0019$ & $0.0944 \pm 0.0009$ \\
    NNclass & $0.0213 \pm 0.0015$ & $0.0362 \pm 0.0015$ & $0.0323 \pm 0.0006$ & $0.0249 \pm 0.0004$ \\
    NF & $\mathbf{0.0125} \pm 0.0019$ & $\mathbf{0.0100} \pm 0.0011$ & $\mathbf{0.0136} \pm 0.0015$ & $\mathbf{0.0147} \pm 0.0019$ \\
\hline                                   %inserts single line
\end{tabular}
\end{table*}

\begin{table*}
\caption{1D Wasserstein distance computed for each galaxy property. 1D scores are computed with the complete test set catalog, and 1000 samples per instance to calculate the standard deviation.}             % title of Table
\label{table:wd1d}      % is used to refer this table in the text
\centering                          % used for centering table

\begin{tabular}{l c c c c}        % centered columns (4 columns)
\hline\hline                 % inserts double horizontal lines
1D Wasserstein & $M_*$ & $g - i$ & sSFR & $R_{1/2}^{(*)}$\\    % table heading 
\hline                        % inserts single horizontal line
    NNgauss & $0.0037 \pm 0.0002$ & $0.0123 \pm 0.0004$ & $0.0377 \pm 0.0004$ & $0.0064 \pm 0.0003$ \\
    NNclass & $0.0031 \pm 0.0003$ & $\textbf{0.0033} \pm 0.0004$ & $0.0051 \pm 0.0004$ & $0.0054 \pm 0.0003$ \\
    NF & $\mathbf{0.0012} \pm 0.0002$ & $0.0042 \pm 0.0009$ & $\mathbf{0.0034} \pm 0.0006$ & $\mathbf{0.0017} \pm 0.0003$ \\
\hline                                   %inserts single line
\end{tabular}
\end{table*}

\begin{table*}
\caption{2D Wasserstein distance computed for each pair of galaxy properties. 2D scores are computed using a randomly drawn subset of $3\times10^4$ instances from the test set, and 5 samples per instance to compute the standard deviation.}             % title of Table
\label{table:wd2d}      % is used to refer this table in the text
\centering                          % used for centering table

\begin{tabular}{l c c c c}        % centered columns (4 columns)
\hline\hline                 % inserts double horizontal lines
2D Wasserstein & $M_* \times g - i$ & $M_* \times $sSFR & $M_* \times R_{1/2}^{(*)}$ & $g - i \times $sSFR\\    % table heading 
\hline                        % inserts single horizontal line
    NNgauss & $0.0193 \pm 0.0012$ & $0.0334 \pm 0.0020$ & $0.0062 \pm 0.0004$ & $0.0399 \pm 0.0021$ \\
    NNclass & $0.0057 \pm 0.0001$ & $0.0069 \pm 0.0002$ & $0.0064 \pm 0.0002$ & $0.0071 \pm 0.0001$ \\
    NF & $\mathbf{0.0051} \pm 0.0007$ & $\mathbf{0.0054} \pm 0.0007$ & $\mathbf{0.0033} \pm 0.0003$ & $\mathbf{0.0067} \pm 0.0012$ \\
\hline                                   %inserts single line
\end{tabular}
\end{table*}

\end{appendix}

% \begin{thebibliography}{}

%   \bibitem[Baker(1966)]{baker} Baker, N. 1966,
%       in Stellar Evolution,
%       ed.\ R. F. Stein,\& A. G. W. Cameron
%       (Plenum, New York) 333

%    \bibitem[Balluch(1988)]{balluch} Balluch, M. 1988,
%       A\&A, 200, 58

%    \bibitem[Cox(1980)]{cox} Cox, J. P. 1980,
%       Theory of Stellar Pulsation
%       (Princeton University Press, Princeton) 165

%    \bibitem[Cox(1969)]{cox69} Cox, A. N.,\& Stewart, J. N. 1969,
%       Academia Nauk, Scientific Information 15, 1

%    \bibitem[Mizuno(1980)]{mizuno} Mizuno H. 1980,
%       Prog. Theor. Phys., 64, 544
   
%    \bibitem[Tscharnuter(1987)]{tscharnuter} Tscharnuter W. M. 1987,
%       A\&A, 188, 55
  
%    \bibitem[Terlevich(1992)]{terlevich} Terlevich, R. 1992, in ASP Conf. Ser. 31, 
%       Relationships between Active Galactic Nuclei and Starburst Galaxies, 
%       ed. A. V. Filippenko, 13

%    \bibitem[Yorke(1980a)]{yorke80a} Yorke, H. W. 1980a,
%       A\&A, 86, 286

%    \bibitem[Zheng(1997)]{zheng} Zheng, W., Davidsen, A. F., Tytler, D. \& Kriss, G. A.
%       1997, preprint
% \end{thebibliography}

\end{document}